\documentclass[amsmath,twocolumn]{revtex4}
\usepackage{graphicx}
\newcommand{\eq}[1]{(\ref{eq:#1})}
\newcommand{\eqname}[1]{\label{eq:#1}}
\newcommand{\kk}{ {\bf k}}
\newcommand{\vv}{ {\bf v}}
\newcommand{\xx}{ {\bf x}}

\newcommand{\Psihd}{\hat\Psi^\dagger}

\newcommand{\Psih}{\hat\Psi}

\newcommand{\Hamilt}{{\mathcal H}}

\begin{document}

\title{Quantum fluid effects and parametric instabilities in microcavities}

\author{Cristiano Ciuti}
\email{ciuti@lpa.ens.fr} \affiliation{Laboratoire Pierre Aigrain,
\'Ecole Normale Sup\'erieure, 24 rue Lhomond, 75005 Paris, France}
\author{Iacopo Carusotto}
\affiliation{CRS BEC-INFM and Dipartimento di Fisica, Universit\`a
di
  Trento, I-38050 Povo, Italy}

%\subjclass[pacs]{71.36.+c, 42.65.-k, 03.75.Kk}
\date{\today}

% Pacs (for editors)
% 71.36.+c Polaritons (including photon-phonon and photon-magnon
%interactions)
%42.50.-p Quantum optics
%42.65.-k Nonlinear optics
%71.35.Lk Collective effects (Bose effects, phase space filling,
% and excitonic phase transitions)
% 03.75.Kk Dynamic properties of
% condensates; collective and hydrodynamic excitations, superfluid
% flow

\begin{abstract}
We present a description of the non-equilibrium properties of a
microcavity polariton fluid, injected by a nearly-resonant
continuous wave pump laser. In the first part, we point out the
interplay between the peculiar dispersion of the Bogolubov-like
polariton excitations and the onset of polariton parametric
instabilities. We show how collective excitation spectra having no
counterpart in equilibrium systems can be observed by tuning the
excitation angle and frequency. In the second part, we explain the
impact of these collective excitations on the in-plane propagation
of the polariton fluid. We show that the resonant Rayleigh
scattering induced by artificial or natural defects is a very
sensitive tool to show fascinating effects such as polariton
superfluidity or polariton Cherenkov effect. We present a
comprehensive set of predicted far-field and near-field images for
the resonant Rayleigh scattering emission.
\end{abstract}

\maketitle

\section{Preface}

In the last decade, the research group of Professor Marc Ilegems
at EPFL has been working intensively and enthusiastically on the
physics and device applications of artificial photonic systems,
such as semiconductor microcavities and photonic crystals. In this
{\it Festschrift} paper, we are going to present a theory of some
exotic physical properties of coherently excited semiconductor
microstructures in the strong exciton-photon coupling regime. We
hope that the rich phenomenology here described will contribute to
a very pleasant celebration of his 65th birthday.

\section{Introduction}

The behavior of a quantum fluid has played an important role in
many fields of condensed matter and atomic physics, ranging from
superconductors to Helium fluids~\cite{ManyBody} and, during the
last decade, Bose-Einstein condensates of cold trapped
atoms~\cite{AtomicBEC}. One of the most dramatic manifestations of
macroscopic coherence of an interacting many-body system is
superfluidity~\cite{Superfluidity}.

In this paper, we will provide a comprehensive theoretical
analysis of the predicted non-equilibrium propagation properties
of a two-dimensional gas of polaritons in a semiconductor
microcavity in the strong exciton-photon coupling regime
\cite{Weisbuch,Review}. Thanks to their photonic component,
polaritons can be coherently excited by an applied laser field and
detected through the emitted light. Thanks to their excitonic
component, polaritons have strong binary interactions, which have
been shown to produce spectacular and rich polariton amplification
effects through matter-wave stimulated collisions
\cite{Savvidis,Ciuti,offbranch,Saba,Huynh,Kundermann}, as well as
spontaneous parametric
instabilities~\cite{Baumberg,Stevenson,HoudreRRSNLin,Whittaker,
Messin,Dasbach,Dasbach_wire}. Recently, a significant amount of
research has been also focusing on the quantum optical properties
of the polariton emission in the parametric regime with the
possibility of observing polariton squeezing and polariton pair
entanglement
\cite{Ciuti_01,squeezing,Schwendimann,twin,Ciuti_branch,Langbein,Langbein_eight,Savasta_entangled}.

Here, we are going to present a detailed discussion of the
interplay between the peculiar polariton collective excitations
and the rich variety of parametric instabilities, which occur in
presence of a nearly-resonant continuous wave pump laser. In
addition, we are going to discuss the impact on the propagation
properties of a moving polariton fluid and analyze in detail the
superfluid regime, which we predicted in a very recent Letter
\cite{PRL_superfluid}. As our system is a strongly non-equilibrium
one, the polariton field oscillation frequency is not fixed by any
equation of state relating the chemical potential to the polariton
density, but it can be tuned by the frequency of the exciting pump
laser. This opens the possibility of having a collective
excitation spectrum which has no counterpart in usual systems
close to thermal equilibrium. In particular, we will analyze the
propagation in presence of a static potential (either for the
photonic or exciton component), which is known to produce resonant
Rayleigh scattering (RRS) of the exciting laser
field~\cite{RRS,HoudreRRSLin,HoudreRRSNLin,Langbein_ring}.
Superfluidity of the polariton fluid manifests itself as a
dramatic collapse of the resonant Rayleigh scattering intensity
when the flow velocity imprinted by the exciting laser beam is
slower than the interaction-induced sound velocity in the
polariton fluid. Furthermore, a dramatic reshaping of the RRS
pattern  due to polariton-polariton interactions can be observed
in both momentum and real space even at higher flow velocities,
e.g. with the appearance of Cherenkov-like patterns. We will
present a rich set of predicted far-field and near-field images of
the resonant Rayleigh scattering emission.

\section{Hamiltonian and polariton mean-field equations}
In order to describe a planar microcavity containing a quantum
well with an excitonic resonance strongly coupled to a cavity
mode, we will consider the following model Hamiltonian
\cite{Ciuti_Review}:
\begin{multline}
  \label{eq:Hamilt_tot}
  \Hamilt=\int\!d\xx\,\sum_{ij=\{X,C\}}
\Psihd_{i}(\xx) \,\Big[{\mathbf
  h}^{0}_{ij}+V_{i}(\xx)\,\delta_{ij} \Big]\,\Psih_{j}(\xx)\\
%+\sum_{i} V_{i}(\xx)\,\Psihd_{i}(\xx)\,\Psih_{i}(\xx) \\
+\frac{\hbar g}{2}\int\! d\xx\,\Psihd_{X}(\xx)\,\Psihd_{X}(\xx)\,
\Psih_{X}(\xx)\,\Psih_{X}(\xx)+\\
+\int\!d\xx\,\hbar F_{p}(\xx,t) \,\Psihd_{C}(\xx)+\textrm{h.c.}~,
\end{multline}
where $\xx$ is the in-plane spatial position and the field
operators $\Psi_{X,C}(\xx)$ respectively describe excitons ($X$)
and cavity photons ($C$). They satisfy Bose commutation rules,
$[\Psih_i(\xx),\Psihd_j(\xx')]=\delta^2(\xx-\xx')\,\delta_{ij}$.
Note that, for simplicity, we will limit our treatment to the case
of polariton modes with the same circular polarization, which can
be excited by a circularly polarized pump. The approach here
presented can be generalized to the spin-dependent case by
considering appropriate spin-dependent exciton-exciton collisional
potentials.

\begin{figure}[t!]
\begin{center}
\includegraphics[width=\columnwidth,clip]{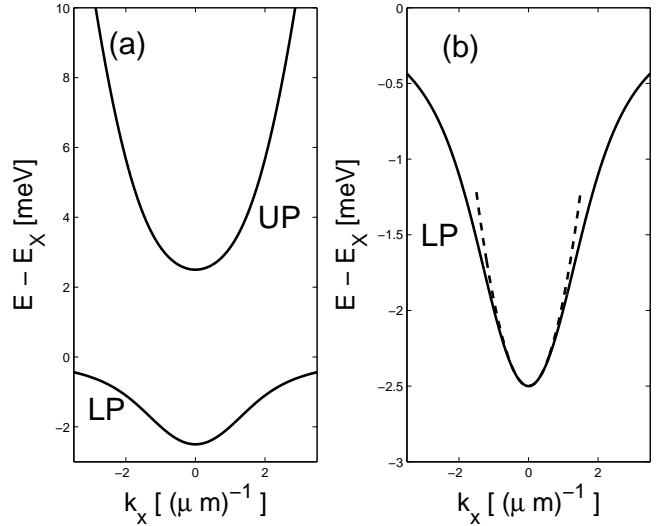}
\caption{(a): Linear dispersion of the Lower (LP) and Upper (UP)
Polariton branches as a function of $k_x$ ($k_y = 0$). Cavity
parameters: $\hbar \omega_X= \hbar \omega_C =1.4\,\textrm{eV}$,
$2\,\hbar \Omega_R=5\,\textrm{meV}$. (b) Zoom of the LP branch.
The dashed line depicts the parabolic approximation around the
bottom of the dispersion. \label{fig:figura1}}
\end{center}
\end{figure}

  The single-particle Hamiltonian ${\mathbf h}^0$ reads
\begin{equation}
  \label{eq:Hamilt_lin}
  {\mathbf h}^0=
\hbar \left(
\begin{array}{cc}
\omega_{X}(-i\nabla) & \Omega_R \\
\Omega_R & \omega_C(-i\nabla)
\end{array}
\right),
\end{equation}
where $\omega_{C}(\kk)=\omega_{C}^0\,\sqrt{1+{\kk^2}/{k_z^2}}$ is
the cavity mode energy dispersion as a function of the in-plane
wavevector $\kk$ and $k_z$ is the quantized photon wavevector
along the growth direction. $\Omega_R$ is the Rabi frequency of
the exciton-cavity photon coupling. In the following, we will
consider a flat exciton dispersion $\omega_X(\kk)=\omega_X$. In
this framework of coupled harmonic oscillators, polaritons simply
arise as the eigenstates of the linear Hamiltonian
\eq{Hamilt_lin}. $\omega_{LP(UP)}(\kk)$ denotes the dispersion of
the lower (upper) polariton branch [Fig.\ref{fig:figura1}(a)].

The term proportional to $F_{p}(\xx,t)$ in Eq.
(\ref{eq:Hamilt_tot}) represents an applied coherent laser pump
spot, which drives the cavity and injects polaritons. In the
following, we will consider the case of a monochromatic laser
field of frequency $\omega_{p}$ and plane-wave profile with
in-plane wave-vector ${\bf k}_p$. The in-plane wave-vector is
linked to the incident direction by the simple relationship
$k_{p}=\sin\theta_p\; \omega_{p}/c$, $\theta_p$ being the pump
incidence angle with respect to the growth direction. This means
that an oblique pump incidence generates a polariton fluid with a
non-zero flow velocity along the cavity plane. For $\hbar \omega_p
= 1400$ meV, an in-plane wavevector $k_p = 1~(\mu$m$)^{-1}$
corresponds to a pump incidence angle of $8.1^{\circ}$.

The nonlinear interaction term in Eq. (\ref{eq:Hamilt_tot}) is due
to exciton-exciton collisional interactions and, as usual, is
modelled by a repulsive ($g>0$) contact potential. The anharmonic
exciton-photon coupling has a negligible effect in the regime
considered in the present study~\cite{Ciuti_Review} and will be
neglected for sake of clarity.

Finally, $V_{X,C}(\xx)$ are the single particle potentials acting
on the exciton and photon fields respectively. These potentials
break the translational symmetry along the cavity plane. The
exciton potential $V_{X}(\xx)$ can be due to natural interface or
alloy disorder in the semiconductor quantum wells due to
unavoidable growth imperfections. The photonic potential
$V_{C}(\xx)$ can be due to fluctuations of the cavity length or
imperfections in the Bragg reflectors (photonic
disorder\cite{Langbein}). More interestingly, $V_{C}(\xx)$ can be
designed and created deliberately by means of lithographic
techniques.

Within the mean-field approximation, the time-evolution of the
mean fields $\psi_{X,C}(\xx)=\langle\Psih_{X,C}(\xx) \rangle$
under the Hamiltonian \eq{Hamilt_tot} is given by:
\begin{widetext}
\begin{equation}
  \label{eq:GPE}
  i\,\frac{d}{dt}
\left(
  \begin{array}{c}
\psi_{X}(\xx) \\ \psi_{C}(\xx)
  \end{array}
\right)= \left(
  \begin{array}{c}
0 \\ F_{p} \, e^{i(\kk_p\xx-\omega_p t)}
  \end{array}
\right)+ \left[ {\mathbf h}^0 + \left(
\begin{array}{cc}
V_X(\xx)+g|\psi_X(\xx)|^2 - i \gamma_X& 0 \\ 0 & V_C(\xx) -i
\gamma_C
  \end{array}
\right) \right] \left(
  \begin{array}{c}
\psi_{X}(\xx) \\ \psi_{C}(\xx)
  \end{array}
\right) .
\end{equation}
\end{widetext}

Using the language of the quantum fluid community, these are the
Gross-Pitaevskii equations~\cite{AtomicBEC} for our
cavity-polariton system. The quantities $\gamma_X$ and $\gamma_C$
represent the homogoneous broadening of the exciton and photon
modes respectively.  In the present work, we will be concerned
with an excitation close to the bottom of the LP dispersion, i.e.
the region most protected~\cite{Review} from the exciton
reservoir, which may be otherwise responsible for
excitation-induced decoherence~\cite{savasta}.

%\section{The stationary state: bistability effects}

\section{Stationary solutions in the homogeneous case}
\begin{figure}[t!]
\begin{center}
\includegraphics[width=\columnwidth,clip]{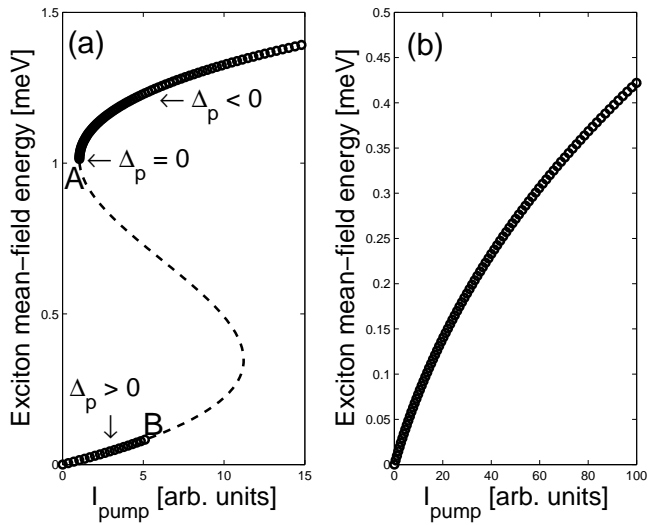}
\caption{Exciton mean-field energy $ \hbar g|\Psi^{ss}_X|^2$ (meV)
as a function of the incident pump intensity (arb. units). Cavity
parameters: $\hbar \gamma_C = \hbar \gamma_X = 0.1\,\textrm{meV}$,
$\hbar \omega_X= \hbar \omega_C^0= 1.4\,\textrm{eV}$, $2\, \hbar
\Omega_R=5\,\textrm{meV}$. (a) Bistability curve obtained with the
excitation parameters: $k_p=0.314\ \mu\textrm{m}^{-1}$ (well in
the parabolic region near the bottom of the LP dispersion), $\hbar
\omega_p-  \hbar \omega_{LP}(\kk_p) =0.47\,\textrm{meV}$. Circles
depict the calculated stable points, while the dashed line
represents the unstable branch. The threshold points A and B are,
respectively, due to a single-mode (Kerr) or a multi-mode
(parametric) instability. (b) Optical limiter curve obtained with
the same $k_p$, but with $\hbar \omega_p-\hbar \omega_{LP}(\kk_p)=
- 0.47\,\textrm{meV}$. In this case, all stationary solutions are
stable. \label{fig_bistability_0314}}.
\end{center}
\end{figure}
In the homogeneous case (i.e., $V_{X,C}(\xx) = 0$ and
translational invariance along the plane), we can look for
spatially homogeneous stationary states of the system in which the
field has the same plane-wave structure
$\psi_{X,C}(\xx,t)=\exp[i(\kk_{p} \xx-\omega_{p}
t)]\,\psi^{ss}_{X,C}$ as the incident laser pump field. The
mean-field equations
\begin{eqnarray}
\Big(\omega_{X}(\kk_{p})-\omega_{p}-\frac{i}{2}\gamma_X+
g\,|\psi_X^{ss}|^2\Big)\,
\psi^{ss}_{X}+\Omega_R\,\psi_C^{ss}=0  \label{eq:ss_X} \\
\Big(\omega_{C}(\kk_{p})-\omega_{p}-\frac{i}{2} \gamma_C \Big)\,
\psi^{ss}_{C}+\Omega_R\,\psi_X^{ss}=-F_p,  \label{eq:ss_C}
\end{eqnarray}
are the non-equilibrium analogous of the state equation, which in
equilibrium systems links the chemical potential to the particle
density. Importantly, we stress that while the oscillation
frequency of the condensate wavefunction in an isolated gas is
equal to the chemical potential $\mu$ and therefore it is fixed by
the equation of state, in the present driven-dissipative system it
is equal to the frequency $\omega_{p}$ of the driving laser and
therefore it is an experimentally tunable parameter. Hence, the
microcavity polariton system allows us to explore a regime, which
is not accessible in systems close to thermal equilibrium, such as
the ultracold trapped atoms.

\section{Linearized Bogoliubov-like theory}

As usual in the theory of nonlinear systems, stability of the
solutions of Eqs. (\ref{eq:ss_X}-\ref{eq:ss_C}) with respect to
fluctuations has to be checked by linearizing Eq. \eq{GPE} around
the stationary state. Perturbations can be produced by classical
fluctuations of the pump field, quantum noise of the exciton and
photon fields as well as the presence of perturbing potentials
$V_{C,X}(\xx)$, which have not been considered by the plane-wave
solutions in Eqs. (\ref{eq:ss_X}-\ref{eq:ss_C}).

In the stability region, the linearized response of the system to
a weak perturbation is analogous to the celebrated Bogoliubov
theory of the weakly interacting Bose gas~\cite{AtomicBEC}.  Let
us define the slowly varying fields with respect to the pump
frequency as
\begin{equation}
\delta\phi_{i}(\xx,t) = \delta {\psi_{i}}(\xx,t) \exp(i\omega_p
t)~,
\end{equation}
and let us consider the four-component displacement vector
\begin{equation}
\delta {\vec \phi}(\xx,t)= \big( \delta\phi_{X}(\xx,t),
\delta\phi_{C}(\xx,t), \delta\phi^*_{X}(\xx,t),
\delta\phi_{C}^*(\xx,t) \big)^T~.
\end{equation}
The equation of motion for the four-component displacement vector
reads
\begin{equation}
  \label{eq:Bogo_motion}
i\frac{d}{dt} \delta{\vec \phi}={\mathcal L}\cdot \delta{\vec
\phi}+{\vec f}_{pert},
 \end{equation}
where ${\vec f}_{pert}$ is the inhomogeneous source term produced
by the perturbation. The expression for ${\vec f}_{pert}$ depends
on which kind of perturbation is considered and will be given
explicitly later for the case of a perturbation induced by the
single particle potentials $V_{C,X}(\xx)$. The linear operator
${\mathcal L}$ is
\begin{widetext}
\begin{equation}
  \label{eq:Bogo_L}
  {\mathcal L}=
\left(
\begin{array}{cccc}
\omega_X+2g\,|\psi_X^{ss}|^2-  \omega_p -\frac{i\gamma_X}{2} &
\Omega_R & g\,\psi^{ss\,2}_X\,e^{2 i \kk_p\xx} & 0 \\
\Omega_R & \omega_C(-i\nabla)-  \omega_p - \frac{i\gamma_C}{2} & 0 & 0 \\
-g\,\psi^{ss\,*\,2}_{X}\,e^{-2 i \kk_p\xx} & 0 &
-\big(\omega_X+2g\,|\psi_X^{ss}|^2\big) +
\omega_p-\frac{i\gamma_X}{2} &
-\Omega_R \\
0 & 0 & -\Omega_R & -\omega_C(-i\nabla)+
\omega_p-\frac{i\gamma_C}{2}
\end{array}
\right).
\end{equation}
\end{widetext}

\subsection{Stability of the stationary solutions}
\begin{figure}[t!]
\begin{center}
\includegraphics[width=\columnwidth,clip]{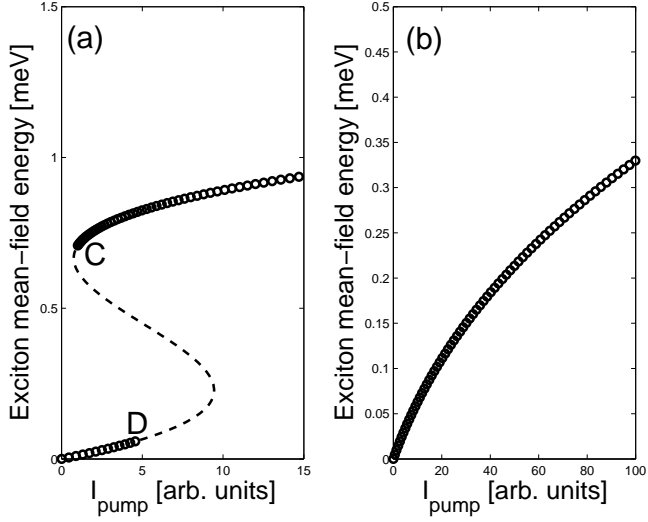}
\caption{Same plot as in Fig. \ref{fig_bistability_0314}, but with
pump in-plane wavevector $k_p=1.5\,\mu\textrm{m}^{-1}$ (close to
the inflection point of the LP dispersion). Pump frequency: (a)
$\hbar \omega_p-  \hbar \omega_{LP}(\kk_p) =0.47\,\textrm{meV}$,
(b) $\hbar \omega_p-  \hbar \omega_{LP}(\kk_p) = -
0.47\,\textrm{meV}$. In contrast to Fig.
\ref{fig_bistability_0314}, here both threshold points C and D are
due to parametric instabilities. \label{fig_bistability_15}}
\end{center}
\end{figure}
The stability of the solutions of Eqs.
(\ref{eq:ss_X}-\ref{eq:ss_C}) can be determined by calculating the
imaginary parts of the eigenvalues of the operator ${\mathcal L}$.
If all the eigenvalues have negative imaginary parts (i.e., as it
happens in the non-interacting case), then the solutions are
stable. Otherwise, an instability occurs. If the polariton
instability involves only the pump mode, we have the analogous of
a {\it Kerr} instability. If the instability is due to pairs of
modes (formation of the so-called signal-idler pairs), we have a
{\it parametric} instability (in the field of quantum fluids, this
kind of dynamical instabilities are generally known as {\em
modulational instabilities}~\cite{NiuModInst}). In Fig.
\ref{fig_bistability_0314}, we have plotted the stationary
solutions for the exciton mean-field energy $\hbar
g\,|\psi^{ss}_X|^2$ (meV) as a function of the incident pump
intensity (arb. units) for realistic microcavity parameters and
with a small pump wave-vector ($k_p=0.314\ \mu\textrm{m}^{-1}$),
close to the bottom of the LP dispersion. In Fig.
\ref{fig_bistability_0314}(a), we have considered the case of a
pump frequency, which is blue-detuned with respect to the
unperturbed lower polariton energy
($\omega_p>\omega_{LP}(\kk_p)$). In this case, there is a clear
S-shaped bistability
curve\cite{BistableExp,BistableExp2,BistableExp3,Bistability}. The
unstable branch, determined through the eigenvalues of the linear
operator ${\mathcal L}$, have been depicted with a dashed line,
while the stable points are represented by circles. The threshold
points A and B are due to a Kerr and to parametric instability
respectively. By comparison, in Fig.
\ref{fig_bistability_0314}(b), we have shown the same quantity,
but for a red-detuned laser frequency
($\omega_p<\omega_{LP}(\kk_p)$). In this case, the polariton
system behaves as an optical limiter\cite{Bistability}, the
absorption is highly sublinear and all the points are stable. To
give a more complete picture, we report in Fig.
\ref{fig_bistability_15}, the analogous calculations, but with a
larger wavevector ($k_p=1.5\ \mu\textrm{m}^{-1}$), close to the
inflection point of the LP dispersion. It is apparent that, while
the shape is analogous, the boundary between the stable and
unstable branches is modified. In particular, the threshold points
C and D are both due to parametric instabilities. Note that nice
hysteresis loops due to polariton bistability have been recently
experimentally demonstrated  in the case $k_p = 0$
\cite{BistableExp} and for a pump wavevector close to the
inflection point of the LP dispersion \cite{BistableExp3}.

\subsection{Complex energy of the collective excitations}
The spectrum of the collective excitations (Bogoliubov modes in
the quantum fluid terminology) can be obtained from the
eigenvalues of the operator ${\mathcal L}$. As the system is
translationally invariant along the plane (we are considering the
homogeneous case $V_X = V_C =0$) , the wavevector $\kk$ is a good
quantum number and the eigenvectors of ${\mathcal L}$ have a
plane-wave form
\begin{equation}
\delta {\vec \phi^{\pm}_{j,{\bf k}}}(\xx)= \left (
\begin{array}{c}
u^{\pm}_{j,X,{\bf k}}~e^{i {\bf k} \xx} \\
u^{\pm}_{j,C,{\bf k}}~e^{i {\bf k} \xx} \\
v^{\pm}_{j,X,{\bf k}} ~e^{i ({\bf k} - 2 \kk_p ) \xx}\\
v^{\pm}_{j,C,{\bf k}}~e^{i ({\bf k} - 2 \kk_p) \xx}
\end{array}
\right )~,
\end{equation}
satisfying the reduced eigenvalue equation
\begin{equation}
\left ( (\omega^{\pm}_{j}(\kk)- \omega_p) {\mathbf 1} -
\tilde{{\mathcal L}}({\bf k},\kk_p) \right ) \cdot \left (
\begin{array}{c}
u^{\pm}_{j,X,{\bf k}} \\
u^{\pm}_{j,C,{\bf k}}  \\
v^{\pm}_{j,X,{\bf k}} \\
v^{\pm}_{j,C,{\bf k}}
\end{array}
\right ) = 0 ~,
\end{equation}
where
\begin{widetext}
\begin{equation}
\label{L_k} \tilde{{\mathcal L}}({\bf k},\kk_p) = \left(
\begin{array}{cccc}
\omega_X+2g\,|\psi_X^{ss}|^2  -\frac{i\gamma_X}{2} &
\Omega_R & g\,\psi^{ss\,2}_X\, & 0 \\
\Omega_R & \omega_C({\bf k})- \frac{i\gamma_C}{2} & 0 & 0 \\
-g\,\psi^{ss\,*\,2}_{X}\, & 0 &
 2 \omega_p -\big(\omega_X+2g\,|\psi_X^{ss}|^2\big) -\frac{i\gamma_X}{2} &
-\Omega_R \\
0 & 0 & -\Omega_R & 2 \omega_p -\omega_C(2 \kk_p - {\bf k})
-\frac{i\gamma_C}{2}
\end{array}
\right).
\end{equation}
\end{widetext}
For each $\kk$, the spectrum is composed by four branches. For
each polariton branch $j\in \{LP,UP\}$, two $\pm$ branches exist,
which are related by the symmetry
\begin{equation}
\omega_j^-(\kk)=2\omega_p-\omega_j^+(2\kk_p-\kk).
\end{equation}

Now, we wish to point out and list clearly the relevant properties
and symmetries in this problem. These properties will be later
discussed in detail through a set of comprehensive examples and
elucidations.

(i) The collective excitations are characterized by the
pump-induced coherent coupling between a generic mode with
wavevector ${\bf k}$ and the "idler" wavevector $2 \kk_p - {\bf
k}$. This corresponds to the elementary process $\{ \kk_p,\kk_p \}
\to \{ {\bf k},2 \kk_p - {\bf k} \}$, i.e. the conversion of two
pump excitations into a signal-idler pair (to use the quantum
optics terminology of parametric oscillators) with the same total
momentum.

(ii) The "idler" branch $\omega^{-}_{LP}(\kk)$ is the "image" of
the ordinary branch $\omega^{+}_{LP}(\kk)$ under the simultaneous
transformations\cite{Ciuti_PL} $\kk\rightarrow 2\kk_{p}-\kk$ and
$\omega\rightarrow 2\omega_{p}-\omega$. The same relationship
holds for the UP branches $\omega^{\pm}_{UP}(\kk)$.

(iii) The matrix $\tilde{{\mathcal L}}({\bf k},\kk_p)$ is
characterized by an {\it anti-hermitian}
 coupling between ${\bf k}$ and $2 \kk_p - {\bf k}$. This
feature is typical of parametric wave-mixing coupling (in the
quantum optics language) or Bogoliubov theory (using the quantum
fluid literature terminology).

(iv) The four branches of eigenvalues $\omega^{\pm}_{UP,LP}(\kk)$
are {\it complex}. As the real parts, the imaginary parts of the
eigenvalues depend both on $\kk$ and on the pump parameters. In
the stability region, all imaginary parts are negative.
\begin{figure}[t!]
\begin{center}
\includegraphics[width=\columnwidth,clip]{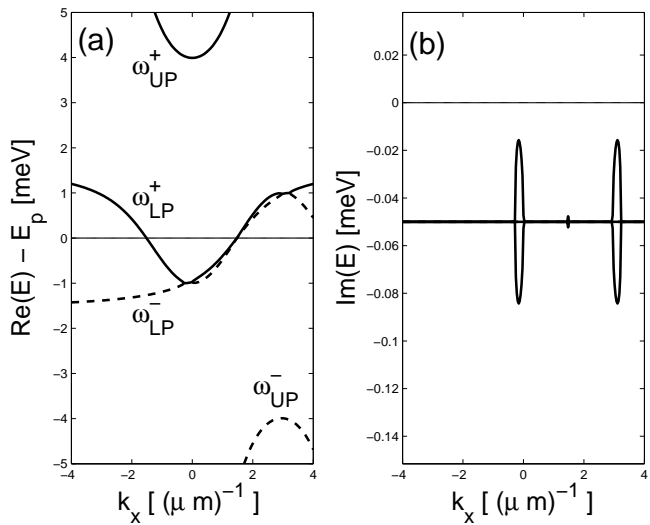}
\caption{(a) Exact energy dispersions $\Re [\hbar
\omega_{LP,UP}^{\pm}]$ of the four polariton Bogoliubov branches
measured with respect to the pump photon energy $\hbar \omega_p$
(meV). (b) Corresponding imaginary parts (meV). Note that negative
imaginary parts imply stability. Excitation parameters: $k_p= 1.5\
\mu\textrm{m}^{-1}$ (along the x-axis), $\hbar \omega_p- \hbar
\omega_{LP}(\kk_p) =0.107\,\textrm{meV}$, $ \hbar g|\Psi^{ss}_X|^2
= 0.05 \textrm{meV}$. Cavity parameters as in the previous
figures. \label{fig_4branches_kx_15}}
\end{center}
\end{figure}

(v) In case of resonant excitation of the lower branch, provided
that the interaction energy $g\,|\psi_{X}^{ss}|^2$ is much smaller
than the polaritonic splitting $\omega_{UP}-\omega_{LP}$, there is
no significant mixing between the LP and UP branches. Hence, a
simplified approach consists in neglecting the contribution of the
upper branch. With this approximation, the branches
$\omega^{\pm}_{LP}(\kk)$ are the eigenvalues of the simplified
matrix
\begin{widetext}
\begin{equation}
\label{L_LP_k} \tilde{{\mathcal L}}_{LP}({\bf k},\kk_p) = \left(
\begin{array}{cc}
\omega_{LP}({\bf k}) +2g_{LP}\,|\psi_{LP}^{ss}|^2
-\frac{i\gamma_{LP}({\bf k})}{2} & g_{LP}\,\psi^{ss\,2}_{LP}\ \\
-g_{LP}\,\psi^{ss\,*\,2}_{LP}\, &
 2 \omega_p -\big(\omega_{LP}(2 \kk_p - {\bf k})+2g_{LP}\,|\psi_{LP}^{ss}|^2\big) -\frac{i\gamma_{LP}({\bf k})}{2} \\
\end{array}
\right)~,
\end{equation}
\end{widetext}
where the stationary lower polariton field is written as a linear
superposition of the exciton and cavity photon fields, namely
\begin{equation}
\label{psi_{LP}} \psi_{LP}^{ss} = X_{LP}(\kk_p) \psi_{X}^{ss} +
C_{LP}(\kk_p) \psi_{C}^{ss}~,
\end{equation}
being $|X_{LP}(\kk_p)|^2$ and $|C_{LP}(\kk_p)|^2$ the exciton and
photon fractions of the lower polariton mode with the pump
wavevector. The effective interaction strength
\begin{equation}
\label{g_{LP}}
 g_{LP}=g\,|X_{LP}(\kk_p)|^2 X_{LP}({\bf k})
X_{LP}(2\kk_p-{\bf k})\end{equation} takes into account for the
exciton fraction of the involved lower polariton modes (pump,
signal and idler). The polariton linewidth is given by
$\gamma_{LP}({\bf k}) = |X_{LP}({\bf k})|^2 \gamma_X +
|C_{LP}({\bf k})|^2 \gamma_C$.
\begin{figure}[t!]
\begin{center}
\includegraphics[width=\columnwidth,clip]{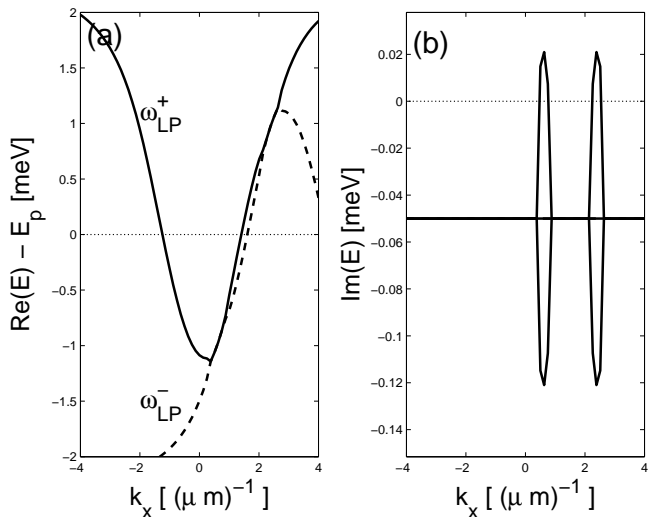}
\caption{(a) Exact energy dispersions $\Re [\hbar
\omega_{LP}^{\pm}]$ of the lower polariton Bogoliubov branches
measured with respect to the pump photon energy $\hbar \omega_p$
(meV). (b) Corresponding imaginary parts (meV). Excitation
parameters: $k_p= 1.5\ \mu\textrm{m}^{-1}$ (along the x-axis),
$\hbar \omega_p- \hbar \omega_{LP}(\kk_p) =0.47\,\textrm{meV}$, $
\hbar g|\Psi^{ss}_X|^2 = 0.699 \textrm{meV}$. Note that here the
stationary solution is unstable, because there are modes with
positive imaginary parts. This unstable point is close to the
point C in Fig. \ref{fig_bistability_15}(a).
\label{fig_2branches_kx_15}}
\end{center}
\end{figure}

(vi) When the diagonal elements of $\tilde{{\mathcal L}}_{LP}({\bf
k},\kk_p)$ are equal, it is easy to verify that $\Re [
\omega^+_{LP}({\bf k)} ] =  \Re [ \omega^-_{LP}({\bf k)} ]$, while
$\Im [ \omega^+_{LP}({\bf k)} ] \neq  \Im [ \omega^-_{LP}({\bf k)}
]$. This means that the parametric coupling produces a splitting
of the imaginary parts of the two LP branches, while the real
parts are the same. If the difference between the diagonal
elements of $\tilde{{\mathcal L}}_{LP}({\bf k},\kk_p)$ is small
compared to the coupling $g_{LP}\,|\psi^{ss}_{LP}|^2$, then the
same property holds. In other words, the dispersions of the two
branches $\omega^+_{LP}({\bf k)}$ and $\omega^-_{LP}({\bf k)}$
stick together, while their imaginary parts are split. One branch
is narrowed with respect to the linear regime, while the other is
overdamped. Analogous properties occur for the exact eigenvalues
of the $4 \times 4$ matrix $\tilde{{\mathcal L}}({\bf k},\kk_p)$,
which are reported in all the figures of this paper.

\begin{figure}[t!]
\begin{center}
\includegraphics[width=\columnwidth,clip]{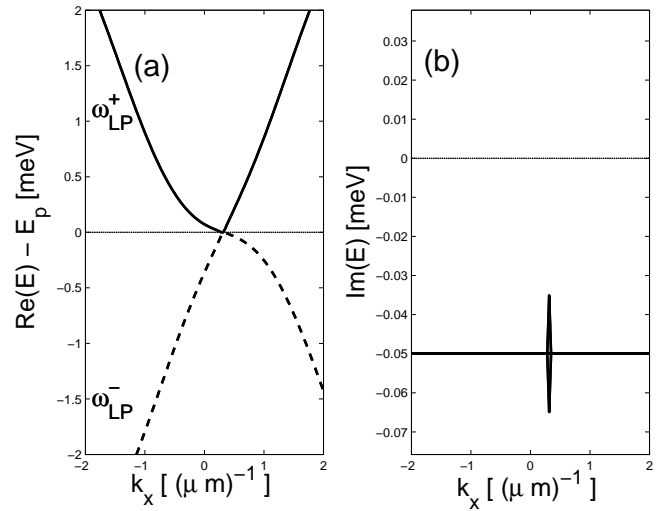}
\caption{(a) Exact energy dispersions $\Re [\hbar
\omega_{LP}^{\pm}]$ of the lower polariton Bogoliubov branches
measured with respect to the pump photon energy $\hbar \omega_p$
(meV). (b) Corresponding imaginary parts (meV). Excitation
parameters: $k_p= 0.314\ \mu\textrm{m}^{-1}$ (along the x-axis),
$\hbar \omega_p- \hbar \omega_{LP}(\kk_p) =0.47\,\textrm{meV}$, $
\hbar g|\Psi^{ss}_X|^2 = 1.02 \textrm{meV}$. Note that this case
is the precursor of a Kerr instability, because the imaginary
parts are modified at the pumped mode only. This stable point is
close to the inversion point A in Fig.
\ref{fig_bistability_0314}(a). \label{fig_2branches_kx_314_Kerr}}
\end{center}
\end{figure}

\subsubsection{Excitation near the inflection point of
the LP dispersion}

In the following, we will show the exact eigenvalues (obtained by
numerical calculations) of the matrix $\tilde{{\mathcal L}}({\bf
k},\kk_p)$ in Eq. (\ref{L_k})  as a function of the excitation
parameters (pump frequency, wavevector and intensity). We will
focus on the subtle interplay between the dramatic modification of
the energy dispersions (depending on the real part of the
eigenvalues) and the onset of the parametric instabilities
(depending on the imaginary part).

As a first example, we consider the case of nearly-resonant
excitation close to the inflection point of the LP dispersion (see
Fig. \ref{fig_4branches_kx_15}). The pump frequency has been taken
slightly blue-detuned with respect to the polariton energy in the
linear regime. In Fig. \ref{fig_4branches_kx_15}(a) the exact
dispersions of the four polariton Bogoliubov branches is shown.
The upper polariton branches are energetically far away and play a
negligible role, while the relevant physics concerns the lower
polariton branches $\omega^{\pm}_{LP}$ only. The corresponding
imaginary parts are shown in \ref{fig_4branches_kx_15}(b). It is
apparent that there is a dramatic modification of the imaginary
part around the wave-vectors $k_x = 0$ and $k_x = 2k_p$. Although
the stationary solutions are here stable (negative imaginary
parts), we can see that we are close to a parametric instability.
In fact, there is one branch, whose imaginary part is not far from
zero. Note that the imaginary parts are split at $\kk_p$ as well,
even if in a much weaker way. This is a precursor of a Kerr (or
single-mode) instability.

In Fig. \ref{fig_2branches_kx_15}, we give another example, with
the same excitation parameters as in Fig. \ref{fig_bistability_15}
(a) and with an exciton mean-field energy $ \hbar g|\Psi^{ss}_X|^2
= 0.699 \textrm{meV}$. In this case, we have an unstable solution,
because, as shown in Fig.\ref{fig_2branches_kx_15}(b), there are
modes with positive imaginary parts. On the bistable curve
 of Fig. \ref{fig_bistability_15}(a), we are here just beyond $C$. The
 instability being of parametric (or multi-mode) type, note that the
 point $C$  is close but does not coincide with the inversion point of
 the bistable curve. In Fig.
\ref{fig_2branches_kx_15}(a), we can see that the branches
$\omega_{LP}^{\pm}$ stick together in the wavevector region where
the parametric instability takes place.

\subsubsection{Excitation near the bottom of the LP
dispersion}

\begin{figure}[t!]
\begin{center}
\includegraphics[width=\columnwidth,clip]{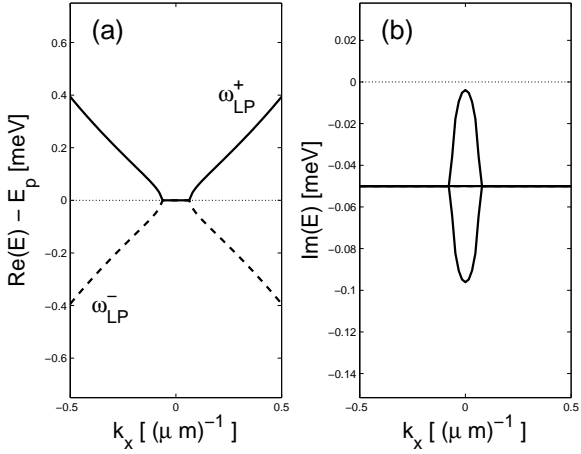}
\caption{(a) Exact energy dispersions $\Re [\hbar
\omega_{LP}^{\pm}]$ of the lower polariton Bogoliubov branches
measured with respect to the pump photon energy $\hbar \omega_p$
(meV). (b) Corresponding imaginary parts (meV). Excitation
parameters: $k_p= 0$, $\hbar \omega_p- \hbar \omega_{LP}(\kk_p)
=0.532\,\textrm{meV}$, $ \hbar g|\Psi^{ss}_X|^2 = 1.2
\textrm{meV}$. \label{fig_2branches_0_plateau.eps}}
\end{center}
\end{figure}
Here, we consider the case of a smaller pump excitation wavevector
and energy , such as to excite the LP branch close to the bottom
of its dispersion. In this region, as shown by Fig.
\ref{fig:figura1}(b), the dispersion of the unperturbed lower
polariton branch is parabolic. In order to stress the non-trivial
effects here predicted, we start by showing a spectacular case,
depicted in Fig.\ref{fig_2branches_kx_314_Kerr}. The excitation
parameters correspond to Fig. \ref{fig_bistability_0314}(a) with
the exciton mean-field energy $\hbar g|\Psi^{ss}_X|^2 = 1.02
\textrm{meV}$, i.e., a point close to the threshold point A for
the Kerr instability. The dispersions of the branches
$\omega_{LP}^{\pm}$ in Fig.\ref{fig_2branches_kx_314_Kerr}(a) have
a corner at the pump wavevector. This dispersion is reminiscent of
the celebrated Bogoliubov linear dispersion in superfluid helium
and in the atomic condensates. If we look at
Fig.\ref{fig_2branches_kx_314_Kerr}(b), we realize that the
modification of the imaginary part is peaked around the pump
wavevector itself. Using the quantum optics language, this is the
precursor of a Kerr instability, because it involves the pumped
mode only.
\begin{figure}[t!]
\begin{center}
\includegraphics[width=\columnwidth,clip]{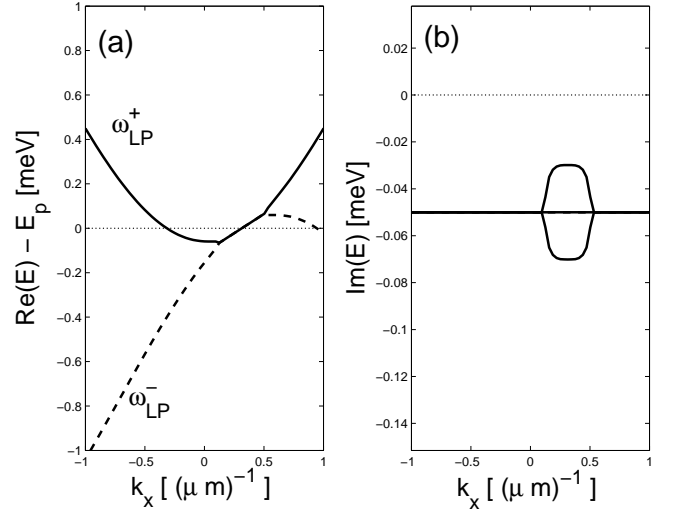}
\caption{(a) Exact energy dispersions $\Re [\hbar
\omega_{LP}^{\pm}]$ of the lower polariton Bogoliubov branches
measured with respect to the pump photon energy $\hbar \omega_p$
(meV). (b) Corresponding imaginary parts (meV). Excitation
parameters: $k_p= 0.314\ \mu\textrm{m}^{-1}$ (along the x-axis),
$\hbar \omega_p- \hbar \omega_{LP}(\kk_p) =0.04\,\textrm{meV}$, $
\hbar g|\Psi^{ss}_X|^2 = 0.04 \textrm{meV}$.
\label{fig_2branches_kx_314_oblique}}
\end{center}
\end{figure}

\begin{figure}[t!]
\begin{center}
\includegraphics[width=\columnwidth,clip]{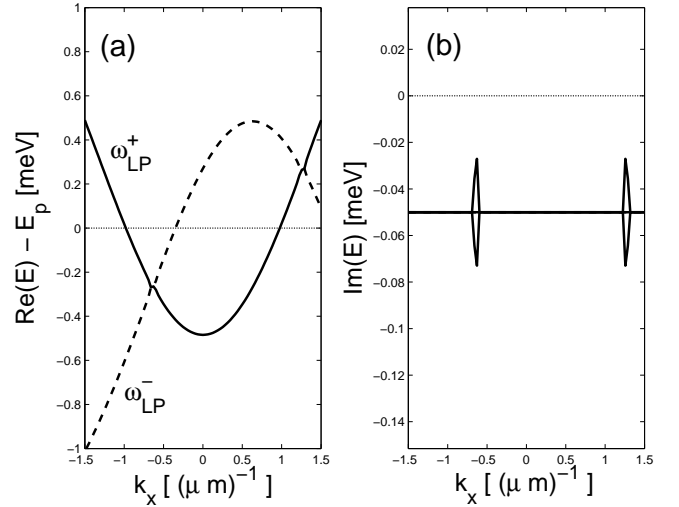}
\caption{(a) Exact energy dispersions $\Re [\hbar
\omega_{LP}^{\pm}]$ of the lower polariton Bogoliubov branches
measured with respect to the pump photon energy $\hbar \omega_p$
(meV). (b) Corresponding imaginary parts (meV). Excitation
parameters: $k_p= 0.314\ \mu\textrm{m}^{-1}$ (along the x-axis),
$\hbar \omega_p- \hbar \omega_{LP}(\kk_p) =0.47\,\textrm{meV}$, $
\hbar g|\Psi^{ss}_X|^2 = 0.04 \textrm{meV}$.
\label{fig_2branches_kx_314_oblique_bis}}
\end{center}
\end{figure}

In Fig. \ref{fig_2branches_0_plateau.eps}, we give another
example, which has no analog in equilibrium systems. Here, we have
a blue-detuned pump at normal incidence, namely $k_p= 0$, $\hbar
\omega_p- \hbar \omega_{LP}(\kk_p) =0.53\,\textrm{meV}$ and $
\hbar g|\Psi^{ss}_X|^2 = 1.2 \textrm{meV}$. In Fig.
\ref{fig_2branches_0_plateau.eps}(a), we can clearly see that the
dispersion of the polariton collective excitations is flat around
the pump wave-vector. For the parameters of Fig.
\ref{fig_2branches_0_plateau.eps}(b), the imaginary parts are all
negative, which implies stability, but, as in
Fig.\ref{fig_2branches_kx_314_Kerr}(b), we are not far from the
onset of a Kerr instability. Note that in Fig.
\ref{fig_2branches_kx_314_oblique}, we have an analogous
situation, but with a finite pump wavevector. As shown by Fig.
\ref{fig_2branches_kx_314_oblique}(a), the branches
$\omega_{LP}^{\pm}$ stick together around the pump wavevector,
with a dispersion exactly linear.

In Fig. \ref{fig_2branches_kx_314_oblique_bis}(a), we show the
dispersions for a stable point, which is close to the threshold
point B in Fig. \ref{fig_bistability_0314}(a). Note that here the
exciton mean-field energy is considerably smaller than the pump
detuning. Hence, the branch sticking occurs in a limited portion
of momentum space, where the imaginary parts are affected, as
shown in Fig. \ref{fig_2branches_kx_314_oblique_bis}(b).

Finally, in Fig. \ref{fig_2branches_kx_314_gap}(a), we give
another different example, with a full gap between the branch
$\omega_{LP}^{+}$ and the branch $\omega_{LP}^{-}$. In Fig.
\ref{fig_2branches_kx_314_gap}(b), we can see that the imaginary
parts of the eigenvalues are unchanged with respect to the linear
regime. Indeed, for these excitation parameters (see the caption
of Fig. \ref{fig_2branches_kx_314_gap}), the microcavity has an
optical limiter behavior similar to the one in Fig.
\ref{fig_bistability_0314}(b) and no instability occurs if the
pump intensity is further increased.

\subsubsection{Simplified analytical model for excitation close to the bottom of
the LP dispersion}

Now, after having shown a few examples of the rich spectra of
non-equilibrium collective excitations and the variety of
interaction-induced polariton instabilities, we present here a
simple approximated approach, which allows us to grasp effectively
the physics contained by the eigenvalues of the matrix in Eq.
(\ref{L_k}). In particular, we consider the case of negligible
mixing with the UP branches, which allows us to focus our analysis
on the simpler $2 \times 2$ matrix in Eq. (\ref{L_LP_k}) instead
of the $4 \times 4$ matrix in Eq. (\ref{L_k}). Moreover, we
consider a pump excitation close to the bottom of the LP
dispersion, where the dispersion is approximately parabolic (see
Fig. \ref{fig:figura1}(b)), i.e.,
\begin{equation}
\omega_{LP}(\kk) \simeq  \omega_{LP}(0) + \frac{\hbar \kk^2}{2
m_{LP}}~,
\end{equation}
where $m_{LP}$ is the effective mass of the LP dispersion. Under
these assumptions, the spectrum of the LP Bogoliubov excitations
can be approximated by the simple expression
\begin{widetext}
\begin{equation}
\omega^\pm_{LP}\simeq \omega_p+ \delta\kk\cdot
\vv_p-\frac{i\gamma_{LP}}{2} \pm\sqrt{(2\,g_{LP}
|\psi_{LP}^{ss}|^2+\eta_{\delta\kk}-\Delta_p)
(\eta_{\delta\kk}-\Delta_p)}, \eqname{Bogo}
\end{equation}
\end{widetext}
where $\delta\kk=\kk-\kk_{p}$,
\begin{equation}
\eta_{\delta\kk}=\frac{\hbar\,\delta\kk^2}{2m_{LP}}~,
\end{equation}
the pump mode flow velocity is $\vv_p=\hbar\kk_p/m_{LP}$ and the
interaction-renormalized pump detuning
\begin{equation}
\label{Delta_p} \Delta_p=\omega_{p}-\omega_{LP}(\kk_p)- g_{LP}
|\psi_{LP}^{ss}|^2.
\end{equation}
In the case of resonant exciton-photon coupling (i.e. $\omega_C(0)
= \omega_{X}$) and small wavevectors, the excitonic fraction of
the lower polariton mode is approximately 0.5. Under these
assumptions, we have $|\psi_{LP}^{ss}|^2 \approx  2
|\psi_{X}^{ss}|^2$ and $g_{LP} \approx g/4$ [see Eqs.
(\ref{psi_{LP}}-\ref{g_{LP}})]. Therefore, $g_{LP}
|\psi_{LP}^{ss}|^2 \approx~0.5~g |\psi_{X}^{ss}|^2$, i.e. the
mean-field interaction energy "felt" by the lower polariton is
half of the mean-field energy for the exciton field. Note that
there are three different cases, according to the value of
$\Delta_p$ defined in Eq. (\ref{Delta_p}).
\begin{figure}[t!]
\begin{center}
\includegraphics[width=\columnwidth,clip]{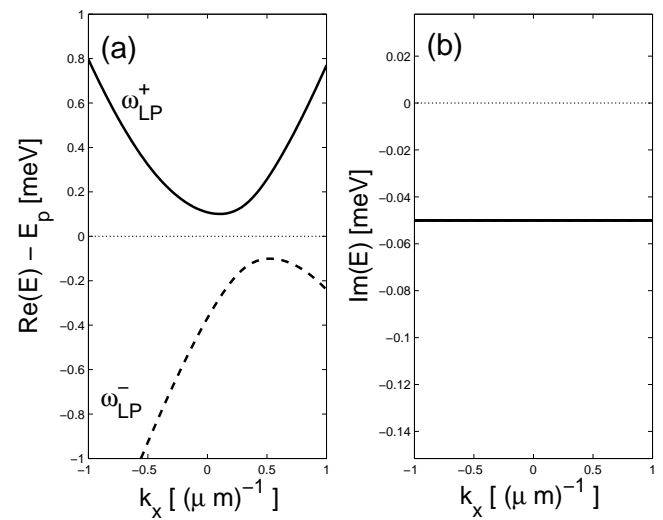}
\caption{(a) Exact energy dispersions $\Re [\hbar
\omega_{LP}^{\pm}]$ of the lower polariton Bogoliubov branches
measured with respect to the pump photon energy $\hbar \omega_p$
(meV). (b) Corresponding imaginary parts (meV). Excitation
parameters: $k_p= 0.314\ \mu\textrm{m}^{-1}$ (along the x-axis),
$\hbar \omega_p- \hbar \omega_{LP}(\kk_p) =0.25\,\textrm{meV}$, $
\hbar g|\Psi^{ss}_X|^2 = 0.6 \textrm{meV}$.
\label{fig_2branches_kx_314_gap}}
\end{center}
\end{figure}

(i) $\Delta_p=0$. In this resonant situation, the $\pm$ branches
touch at $\kk=\kk_{p}$. The effect of the finite flow velocity
$\vv_p$ is to tilt the standard Bogoliubov
dispersion~\cite{AtomicBEC} via the term
  $\delta\kk\cdot\vv_p$.
  While in the non-interacting case the
  dispersion is parabolic, in
  the presence of interactions [Fig.\ref{fig_2branches_kx_314_Kerr}(a)] its
  slope has a
  discontinuity at $\kk=\kk_{p}$. On each side of the corner,
  the $+$ branch starts linearly with group velocities
  respectively given by $v_g^{r,l}=c_s\pm v_p$, $c_s$ being the usual
  sound velocity of the
  interacting Bose gas
  \begin{equation}
c_s=\sqrt{\hbar\ g_{LP} |\Psi^{ss}_{LP}|^2 /m_{LP}}~. \label{c_s}
\end{equation}
On the hysteresis curve of Fig.\ref{fig_bistability_0314}(a), the
condition $\Delta_p=0$ corresponds to the inversion point A.

(ii) $\Delta_p > 0$. In this case, the argument of the square root
in \eq{Bogo} is negative for the wavevectors $\kk$ such that
  $\Delta_p>\eta_{\delta\kk}>\Delta_p-2\, g|\Psi^{ss}_X|^2 $. In this
  region, the $\pm$ branches stick together~\cite{Ciuti_Review}
  (i.e. $\Re[\omega_{LP}^+] = \Re[\omega_{LP}^-]$) and have an
  exactly linear dispersion of slope $\vv_p$ as in Fig. \ref{fig_2branches_kx_314_oblique}(a) and
  in Fig. \ref{fig_2branches_0_plateau.eps}(a). The imaginary parts are instead split, with
  one branch being narrowed and the other broadened~\cite{Ciuti_Review,Ciuti_PL}.
  Increasing further the pump density, the multi-mode parametric
  instability~\cite{Ciuti_PL}  sets in, corresponding to the point B in Fig.
  \ref{fig_bistability_0314}(a).

  (iii) $\Delta_p < 0$.  In this case, as it is shown
in Fig.\ref{fig_2branches_kx_314_gap}, the branches no longer
touch each other at $\kk_p$ and a full gap between them opens up
for sufficiently large values of $|\Delta_p|$. In Fig.
\ref{fig_bistability_0314}(a), the region $\Delta_p < 0$ is
indicated.

\section{Response to a static potential: Resonant Rayleigh
scattering} The dispersion of the polariton elementary excitations
is the starting point for a study of the microcavity response to a
perturbation. In particular, we shall consider here a moderate
static disorder as described by the potential $V_{C,X}(\xx)$. In
this case the perturbation source term for the equations of the
linearized theory (see Eq. \ref{eq:Bogo_motion}) is the
time-independent quantity
\begin{equation}
{\vec f}_{d}(\xx) = \left (
\begin{array}{c}
V_X(\xx)\,\phi_X^{ss} \\
V_C(\xx)\,\phi_C^{ss} \\
-V_X(\xx)\,\phi_X^{ss\,*} \\
-V_C(\xx)\,\phi_C^{ss\,*}
\end{array}
\right ) ~.
\end{equation}
The induced perturbation of the exciton and photon fields is given
by the expression
\begin{equation}
\delta{\vec \phi}_{d}(\xx) =-{\mathcal
  L}^{-1}\cdot {\vec f}_d(\xx)~.
\label{disorder_x}
\end{equation}

We remind you that, as shown by Eq. (\ref{eq:Bogo_L}), ${\mathcal
  L}$ is an operator depending on the two-dimensional spatial gradient $\nabla$.
It is convenient to perform a spatial Fourier transform, which
leads to the algebraic result
\begin{widetext}
\begin{equation}
\left (
\begin{array}{c}
\delta \tilde{\phi}_X({\bf k}) \\
\delta \tilde{\phi}_C({\bf k})  \\
\delta \tilde{\phi}^{*}_X(2 \kk_p - \kk) \\
\delta \tilde{\phi}^{*}_C(2 \kk_p - \kk)
\end{array}
\right ) = - (\tilde{{\mathcal L}}({\bf k},\kk_p) - \hbar
\omega_p)^{-1} \cdot \left (
\begin{array}{c}
\tilde{V}_X(\kk)\,\phi_X^{ss}\ \\
\tilde{V}_C(\kk)\,\phi_C^{ss}\  \\
- \tilde{V}_X(\kk - 2\kk_p)\,\phi_X^{ss\,*}\ \\
- \tilde{V}_X(\kk- 2 \kk_p)\,\phi_C^{ss\,*}\
\end{array}
\right ) ~, \label{k_eq}
\end{equation}
\end{widetext}
where the eigenvalues of the matrix $\tilde{{\mathcal L}}({\bf
k},\kk_p)$, defined in Eq. (\ref{L_k}), are the 4 branches of
polariton Bogoliubov excitations. The perturbation potentials
$V_{C,X}(\xx)$ break the planar translational symmetry of the
microcavity system, thus exciting polariton modes with in-plane
wavevectors different from the pump wavevector $\kk_p$. However,
being $V_{C,X}(\xx)$ static, the resonant excitation concerns only
Bogoliubov modes whose frequency is equal to $\omega_p$ (within
the polariton homogeneous linewidth). The observable quantity is
\begin{equation}
I_{RRS}({\kk}) \propto |\delta \tilde{\phi}_C({\kk})|^2~,
\end{equation} i.e., the perturbation-induced intensity of the photonic field,
which is proportional to the far-field images of the resonant
Rayleigh scattering signal
\cite{HoudreRRSLin,HoudreRRSNLin,Langbein}.
\begin{figure}[t!]
\begin{center}
\includegraphics[width=7cm]{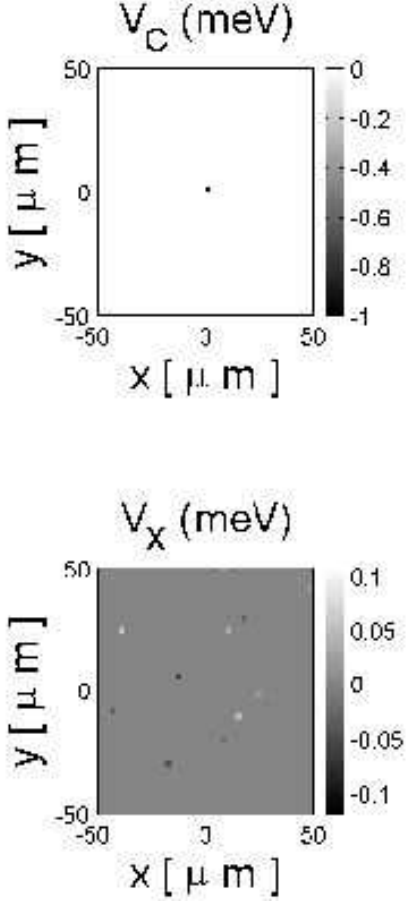}
\caption{Single particle potentials considered in the numerical
calculations. Note that this is a zoom around the origin of the
$400 \mu$m $\times$ $400 \mu$m box (with a $256 \times 256$ grid)
used for the numerical calculations. Top panel: Photonic potential
$V_C(\xx)$ (meV) (it can model an artificial or natural point
defect at the origin $x=y=0$). Bottom panel: Excitonic potential
$V_X(\xx)$ (disordered spatial fluctuations of the exciton
energy). The gray color scale is different with respect to the top
panel. \label{fig_potentials.eps}}
\end{center}
\end{figure}

\subsection{Weak excitation regime and elastic RRS ring}
\begin{figure}[t!]
\begin{center}
\includegraphics[width=6cm,clip]{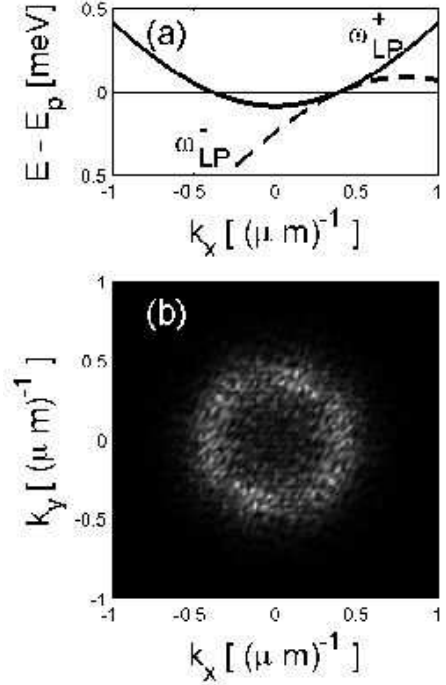}
\caption{ Energy dispersion of the LP branches in the weak
excitation regime.(b) Intensity (arb. units) of the photonic
resonant Rayleigh scattering signal $|\delta
\tilde{\phi}_C(\kk)|^2$. Excitation parameters: $k_p= 0.4 \
\mu\textrm{m}^{-1}$ (along the x-axis), $\hbar \omega_p - \hbar
\omega_{LP}(\kk_p) = 0\,\textrm{meV}$, $ \hbar g|\Psi^{ss}_X|^2 =
0.0001 \textrm{meV}$. The photonic and exciton potentials are
those shown in Fig \ref{fig_potentials.eps}. The speckles of the
elastic RRS ring are due to the disordered excitonic potential.
The chosen plot range is such that a few speckles saturate the
gray scale. \label{fig_linearRRS.eps}}
\end{center}
\end{figure}
\begin{figure}[b!]
\begin{center}
\includegraphics[width=7cm,clip]{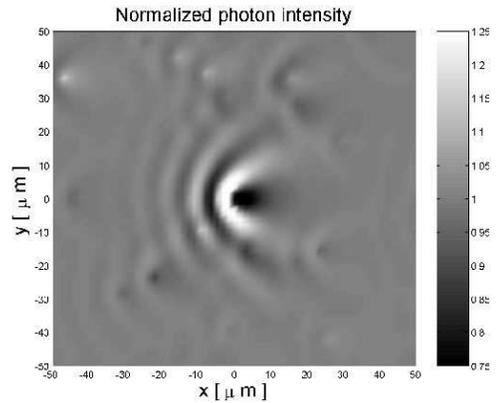}
\caption{ Spatial profile of the normalized cavity photon density,
i.e.,  $I_C(\xx)/I_C^{hom}$. Excitation parameters and potentials
as in Fig. \ref{fig_linearRRS.eps}. The coherent diffusion pattern
induced by the point defect at the origin is the main feature on
top of the random landscape produced by the exciton disorder.
\label{fig_linearRRS_space.eps}}
\end{center}
\end{figure}
\begin{figure}[t!]
\begin{center}
\includegraphics[width=6cm,clip]{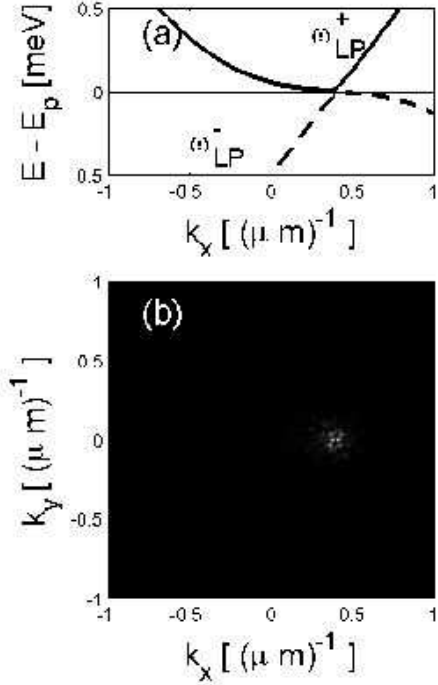}
\caption{Superfluid regime. Same parameters as in Fig.
\ref{fig_linearRRS.eps}, but with $\hbar \omega_p - \hbar
\omega_{LP}(\kk_p) = 0.467\,\textrm{meV}$, $ \hbar
g|\Psi^{ss}_X|^2 = 1 \textrm{meV}$. In this superfluid regime, the
RRS elastic ring has collapsed. \label{fig_superfluidRRS.eps}}
\end{center}
\end{figure}
\begin{figure}[h]
\begin{center}
\includegraphics[width=7cm,clip]{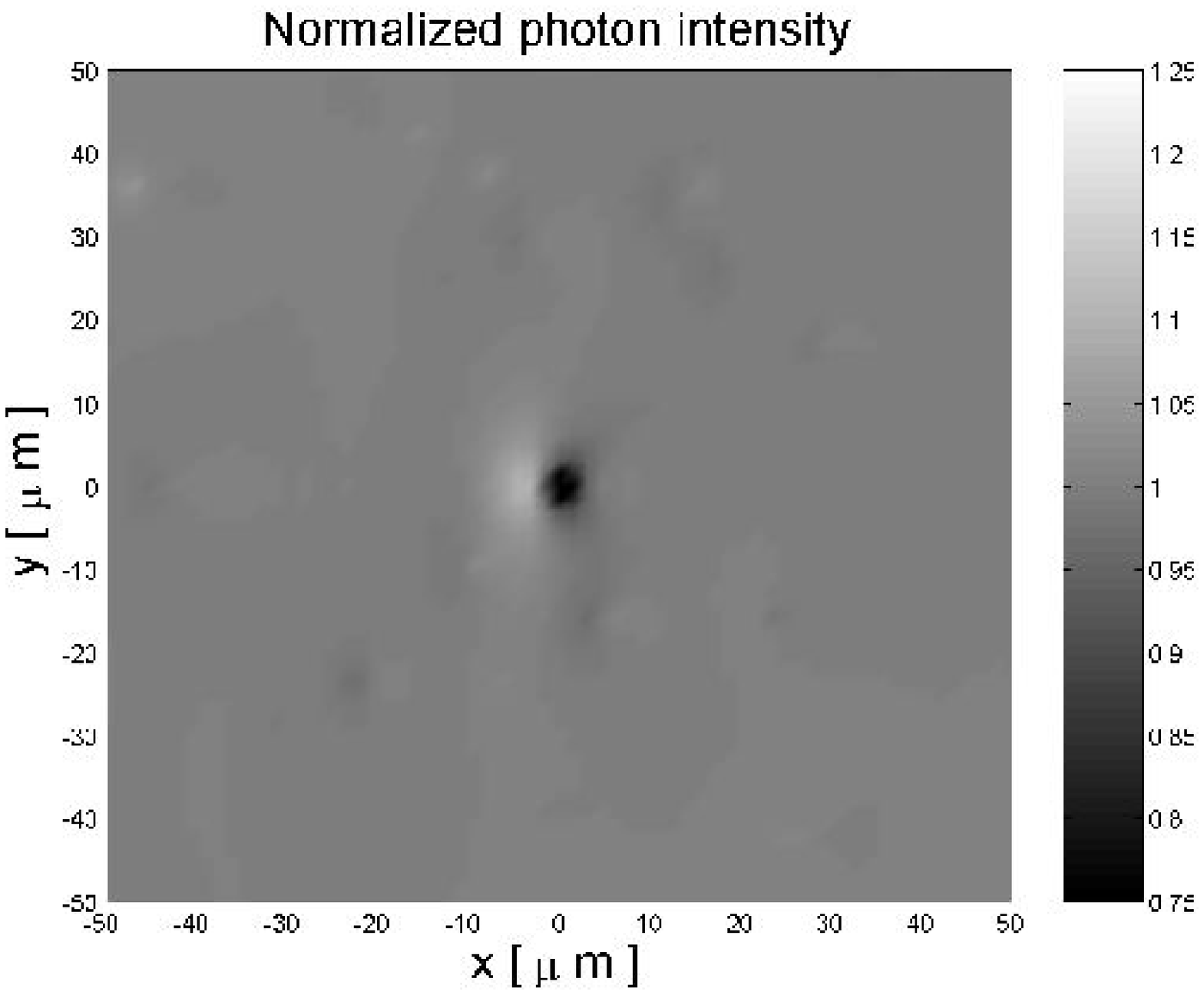}
\caption{ Spatial profile of the normalized cavity photon density
in the superfluid regime. Excitation parameters as in Fig.
\ref{fig_superfluidRRS.eps}. To compare with the normal (weak)
excitation regime, see Fig. \ref{fig_linearRRS_space.eps}.
\label{fig_superfluidRRS_space.eps}}
\end{center}
\end{figure}
In the following, we will show a few applications of Eq.
(\ref{k_eq}), using the perturbation potentials depicted in Fig.
\ref{fig_potentials.eps}. Here, we have considered a single
photonic defect (depth 1 meV, width $1.5 \mu$m) for the in-plane
photonic potential $V_C(\xx)$ [see
Fig.\ref{fig_potentials.eps}(a)] and a disordered excitonic
potential [see Fig.\ref{fig_potentials.eps}(b)]. We point out that
these potentials are just an example and that Eq. (\ref{k_eq}) can
be readily applied for an arbitrary set of perturbation potentials
(see, e.g., the cover of this special volume). In the numerical
applications here reported, we have considered a $256 \times 256$
spatial grid and a squared box ($400 \mu$m $\times$ $400 \mu$m),
with the photonic dot at the center of the box.

In Fig. \ref{fig_linearRRS.eps}, we show the results for the weak
excitation regime, where the many-body effects produced by
polariton-polariton interactions are negligible. In this linear
regime, we have the conventional unperturbed dispersions of the
lower and upper polariton branches. In Fig.
\ref{fig_linearRRS.eps}(a), we have considered the case of
resonant excitation close to the bottom of the LP branch, where
the dispersion is parabolic. In Fig. \ref{fig_linearRRS.eps}(b),
the intensity of the resonant Rayleigh scattering is shown,
displaying the well known elastic ring. In fact, in the linear
regime, the solutions of the equations $\omega_{LP}(\kk) =
\omega_p$ are $\kk$-points on a circle, because the unperturbed
polariton dispersion depends on $|\kk|$ only. The speckles on top
of the elastic ring are due to the random nature of the excitonic
potential.The width of the ring is due to the finite homogeneous
broadening of the polariton modes. Note that, in order to excite
the elastic ring corresponding to the wavevector $\kk_p$, the
Fourier component $\tilde{V}_{C,X}({\kk_p})$ of the static
potentials need to be finite. This condition is easily fulfilled
by a typical excitonic disordered potential or by a photonic
defect whose width is of the order of 1 $\mu$m.

In Fig. \ref{fig_linearRRS_space.eps}, we show the corresponding
spatial pattern. Precisely, we have plotted the normalized
quantity
\begin{equation}
\frac{I_C(\xx)}{I_C^{hom}} =  \frac{|\phi^{ss}_{C} e^{i \kk_p \xx}
+ \delta \phi_C(\xx)|^2}{|\phi^{ss}_{C}|^2}~,
\end{equation}
i.e., the total photon field intensity (homogeneous solution +
potential-induced perturbation) normalized to the intensity of the
homogeneous solution without the potential. For the considered
potentials in Fig. \ref{fig_potentials.eps}, the dominant feature
is due to the photonic point defect. The polariton plane wave
driven by the pump is coherently scattered by the photonic defect
(located at the position $x=y=0$), producing a peculiar
interference pattern, characterized by parabolic wavefronts. In
fact, the polariton field scattered by the point defect is a
cylindrical wave. Hence, if we consider only one defect and
$\kk_p$ is along the x-direction, the total field has the form
$f(\xx) = e^{i k_p x} + \beta e^{i k_p \sqrt{x^2 + y^2}} $. The
constant phase curves are given by the condition $k_p x + k_p
\sqrt{x^2 + y^2} = 2 \pi n$, whose solutions describe parabolic
wavefronts with a symmetry axis oriented along the x-direction, as
nicely depicted by the exact solution in Fig.
\ref{fig_linearRRS_space.eps}. Due to the presence of the exciton
potential, additional disordered features are superimposed on the
main interference pattern produced by the photonic point defect.

\subsection{Superfluid regime}
In presence of interactions, we have seen that the spectrum of
polariton Bogoliubov excitations is dramatically different from
the unperturbed case. This manifestation of polariton many-body
physics can be probed in a sensitive way by the resonant Rayleigh
scattering emission. In Fig. \ref{fig_superfluidRRS.eps}, we start
by considering the most spectacular regime of polariton
superfluidity. This regime can be achieved when the pump is
resonant with the interaction-renormalized polariton dispersion at
the pump wavevector ($\Delta_p = 0$) and when the sound velocity
$c_s$ [see Eq. (\ref{c_s})] of the interacting polariton fluid is
larger than the flow velocity $v_p = \hbar k_p/m_{LP}$ imprinted
by the pump beam. This situation is more favorable to obtain for
excitation close to the bottom of the LP dispersion, implying
smaller pump flow velocity $v_p$ and smaller excitation density
necessary to have $c_s > v_p$. As depicted by Fig.
\ref{fig_superfluidRRS.eps}(a), in this case, the equation $\Re
{[\omega_{LP}^{\pm}(\kk)}] - \omega_p = 0$ has no solutions for
$\kk \neq \kk_p$, meaning that no final states are available for
the elastic scattering induced by the static potential. As a
dramatic consequence, the elastic ring in Fig.
\ref{fig_linearRRS.eps}(b) collapses. As shown in Fig.
\ref{fig_superfluidRRS.eps}(b), only a weak emission around the
pump wavevector $\kk_p$ is left, due to non-resonant processes,
which are allowed by the finite broadening of the polariton modes.

The real space pattern is shown in Fig.
\ref{fig_superfluidRRS_space.eps}, showing that the effect of the
disorder remains localized around the defect positions. Hence, the
polaritonic propagation is {\em superfluid}. In analogous way to
liquid Helium and atomic condensates~\cite{ManyBody,AtomicBEC}, we
can state that the polariton fluid has a superfluid behaviour
according to the Landau criterion, if and only if both following
conditions are satisfied

(a) $\omega_{LP,UP}^{+}(\kk) > \omega_p$ for every $\kk\neq\kk_p$.

(b) $\omega_p > \omega_{LP}(0)$, i.e. there is an elastic ring in
the weak excitation regime.

We point out that the condition (b) is necessary to have a
meaningful definition of polariton superfluidity. In fact, if
$\omega_p < \omega_{LP}(0)$, already in the weak excitation regime
there are no real states at the pump energy and there is no
resonant Rayleigh scattering elastic ring. Note that the
conditions (a) and (b) are achieved not only in the resonant case
$\Delta_p = 0$. Within the parabolic approximation in Eq.
\eq{Bogo}, conditions (a) and (b) are satisfied when $\Delta_p
\leq 0$ {\it and}
\begin{equation}
\label{condition}
 - g_{LP} |\psi_{LP}^{ss}|^2 - \frac{m_{LP}v_p^2}{2 \hbar} < \Delta_p < g_{LP} |\psi_{LP}^{ss}|^2 -
 \frac{m_{LP}v_p^2}{\hbar}~.
\end{equation}

The calculations here reported show the robustness of the
superfluid flow with respect to elastic processes, such as the
scattering on static defects. As the Landau criterion for
superfluidity involves also inelastic processes (e.g., emission of
crystal phonons or heating of residual free carriers), it is
important to note that whenever the superfluidity condition (a)
for  elastic scattering is fulfilled, then it is satisfied {\em a
fortiori} also for the inelastic channels. In fact, the stability
of the mean-field solution implies that almost no Bogoliubov
quasi-particles are present above a stable mean-field solution, so
whenever $\omega_{LP}^+(\kk)>\omega_p$, no final states are
available for the polariton system to lower its energy by
transferring energy to its environment. This is here assumed to be
at almost zero temperature, so that it can only absorb energy from
the polariton system. Friction is therefore absent in this regime.

On the other hand, at a finite temperature of the crystal a
thermally excited "normal" component can appear in the polariton
fluid because of its heating due the interaction with the phonon
bath and the residual free carriers. As it happens in liquid
Helium, this normal component experiences a finite friction when
flowing onto the defect. However, its magnitude vanishes at low
temperatures, and in any case it does not affect the superfluidity
of the co-existing superfluid component in a two-fluid picture.

\subsection{Precursors of parametric instabilities and branch
sticking}

\begin{figure}[t!]
\begin{center}
\includegraphics[width=6cm,clip]{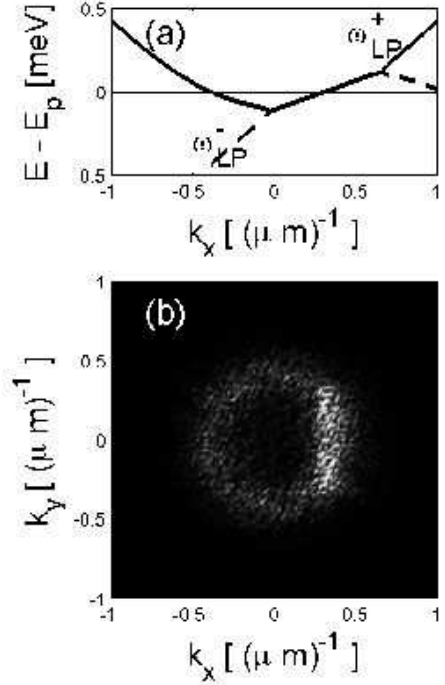}
\caption{Same parameters as in Fig. \ref{fig_linearRRS.eps}, but
with $\hbar \omega_p - \hbar \omega_{LP}(\kk_p) =
0.1\,\textrm{meV}$, $ \hbar g|\Psi^{ss}_X|^2 = 0.07 \textrm{meV}$.
Branch sticking and amplified RRS are precursors of a parametric
instability.
 \label{fig_stickingRRS.eps}}
\end{center}
\end{figure}

\begin{figure}[h]
\begin{center}
\includegraphics[width=7cm,clip]{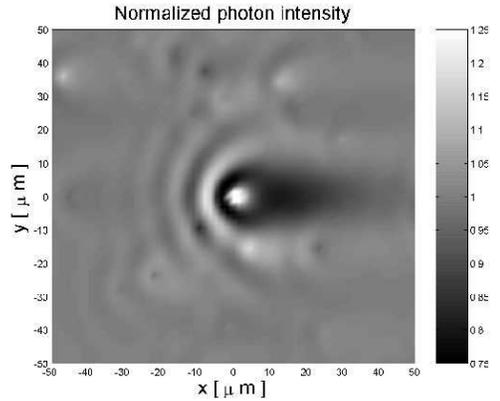}
\caption{ Spatial profile of the normalized cavity photon density.
Excitation parameters as in Fig. \ref{fig_stickingRRS.eps}. Note
that the gray scale of this plot is saturated in the region around
the point defect at the origin. \label{fig_stickingRRS_space.eps}}
\end{center}
\end{figure}

\begin{figure}[h]
\begin{center}
\includegraphics[width=6cm,clip]{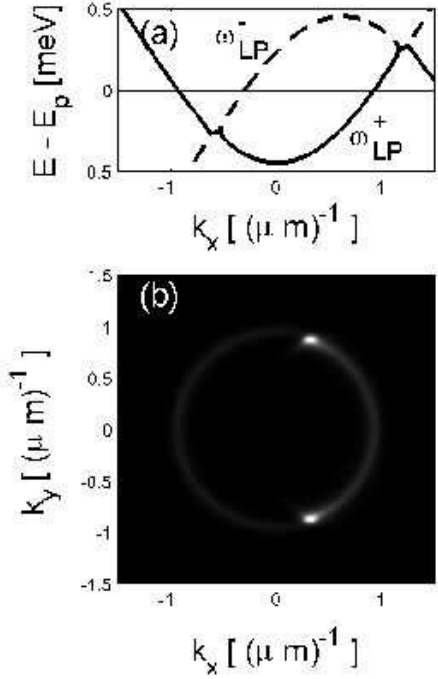}
\caption{Excitation parameters: $k_p= 0.314 \ \mu\textrm{m}^{-1}$
(along the x-axis), $\hbar \omega_p - \hbar \omega_{LP}(\kk_p) =
0.47\,\textrm{meV}$, $ \hbar g|\Psi^{ss}_X|^2 = 0.075
\textrm{meV}$. Note that the gray scale of this plot is saturated
around the two brightest points. Here, we consider the situation
of a dominant photonic potential ($V_X = 0$).
\label{fig_2points.eps}}
\end{center}
\end{figure}
\begin{figure}[t]
\begin{center}
\includegraphics[width=7cm,clip]{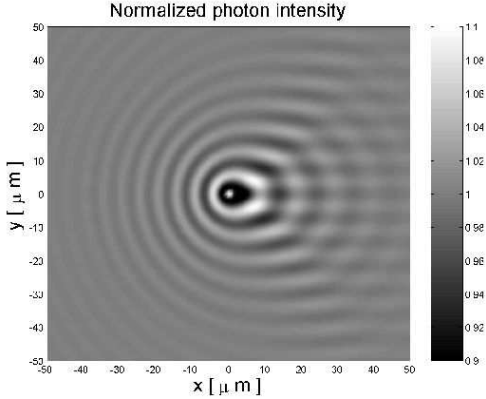}
\caption{Spatial profile of the normalized cavity photon density.
Same parameters as in Fig. \ref{fig_2points.eps}.
\label{fig_2points_space.eps}}
\end{center}
\end{figure}

\begin{figure}[t]
\begin{center}
\includegraphics[width=6cm,clip]{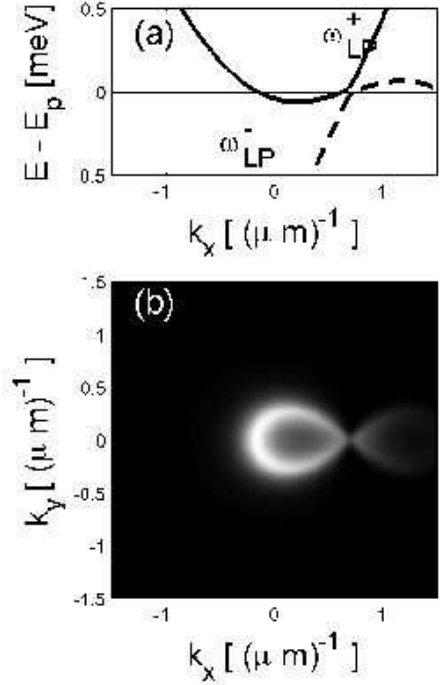}
\caption{Cherenkov regime. Parameters: $\omega_X = \omega_C(0) - 1
\textrm{meV}$,  $k_p= 0.7 \mu\textrm{m}^{-1}$ (along the x-axis),
$\hbar \omega_p - \hbar \omega_{LP}(\kk_p) = 0.599\,\textrm{meV}$,
$ \hbar g|\Psi^{ss}_X|^2 = 1 \textrm{meV}$. Here, we consider the
situation of a dominant photonic potential ($V_X = 0$).
 \label{fig_CherenkovRRS.eps}}
\end{center}
\end{figure}

\begin{figure}[h!]
\begin{center}
\includegraphics[width=7cm,clip]{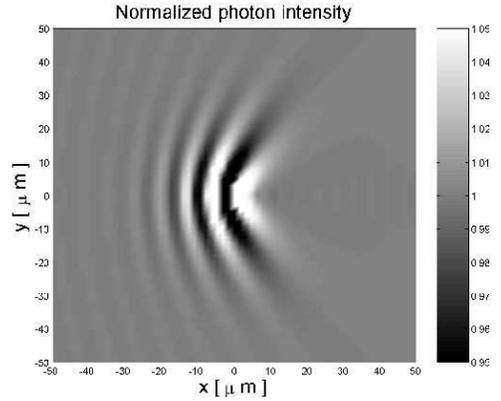}
\caption{ Spatial profile of the normalized cavity photon density.
Same parameters as in Fig. \ref{fig_CherenkovRRS.eps}. The
photonic point defect produces Cherenkov-like wavefronts.
\label{fig_CherenkovRRS_space.eps} Note that, in order to show the
peculiar shape of the wavefronts, the gray scale of this plot is
saturated in the region around the point defect.}
\end{center}
\end{figure}
In the case $\Delta_p > 0$ (i.e., pump frequency higher than the
renormalized frequency of the pumped mode), the resonant Rayleigh
scattering response is completely different, as shown in Fig.
\ref{fig_stickingRRS.eps} and \ref{fig_2points.eps}. In this
regime, the two LP branches stick together, while there is a
splitting of their imaginary parts [see, e.g., the analogous
situation in Fig. \ref{fig_2branches_kx_314_oblique} and Fig.
\ref{fig_2branches_kx_314_oblique_bis}]. Such a scenario
represents the precursor of a multi-mode parametric instability,
which can be triggered by further increasing the excitation
density. In contrast to the superfluid regime, a deformed RRS ring
is apparent in Fig. \ref{fig_stickingRRS.eps}(b) and Fig.
\ref{fig_2points.eps}(b). In particular, depending on the topology
of the $\kk$-space region where the LP branches stick, the RRS
intensity is strongly amplified either on a segment parallel to
$y$ including the point $\kk_p$
[Fig.\ref{fig_stickingRRS.eps}(b)], or around two points of the
straight line parallel to $y$ and passing through the point
$\kk_p$ [Fig. \ref{fig_2points.eps}(b)]. The main consequence of
this in the real-space pattern of Fig.
\ref{fig_stickingRRS_space.eps} is an overall amplification of the
density modulation induced by the defect, in stark contrast with
the superfluid regime. In particular, note the long "shadow" in
the downstream direction with respect to the central defect, which
extends to relatively far distances thanks to the linewidth
narrowing effect. In Fig.\ref{fig_2points_space.eps}, the shadow
of the defect is even more peculiar, showing a series of fringes
parallel to the $x$ direction. These can be explained in terms of
the interference between the pump and the two peaks in $\kk$-space
shown in Fig.\ref{fig_2points.eps}(b).

\begin{figure}[t]
\begin{center}
\includegraphics[width=6cm,clip]{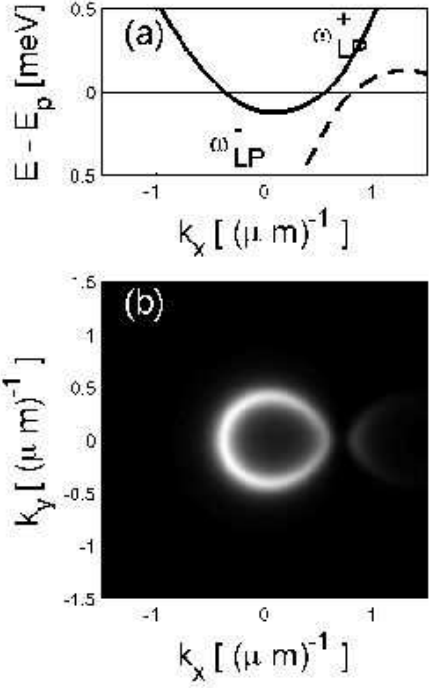}
\caption{Parameters: $\omega_X = \omega_C(0)$, $k_p= 0.7
\mu\textrm{m}^{-1}$ (along the x-axis), $\hbar \omega_p - \hbar
\omega_{LP}(\kk_p) = 0.3\,\textrm{meV}$, $ \hbar g|\Psi^{ss}_X|^2
= 0.6 \textrm{meV}$. \label{fig_gapRRS.eps}}
\end{center}
\end{figure}
\begin{figure}[h]
\begin{center}
\includegraphics[width=7cm,clip]{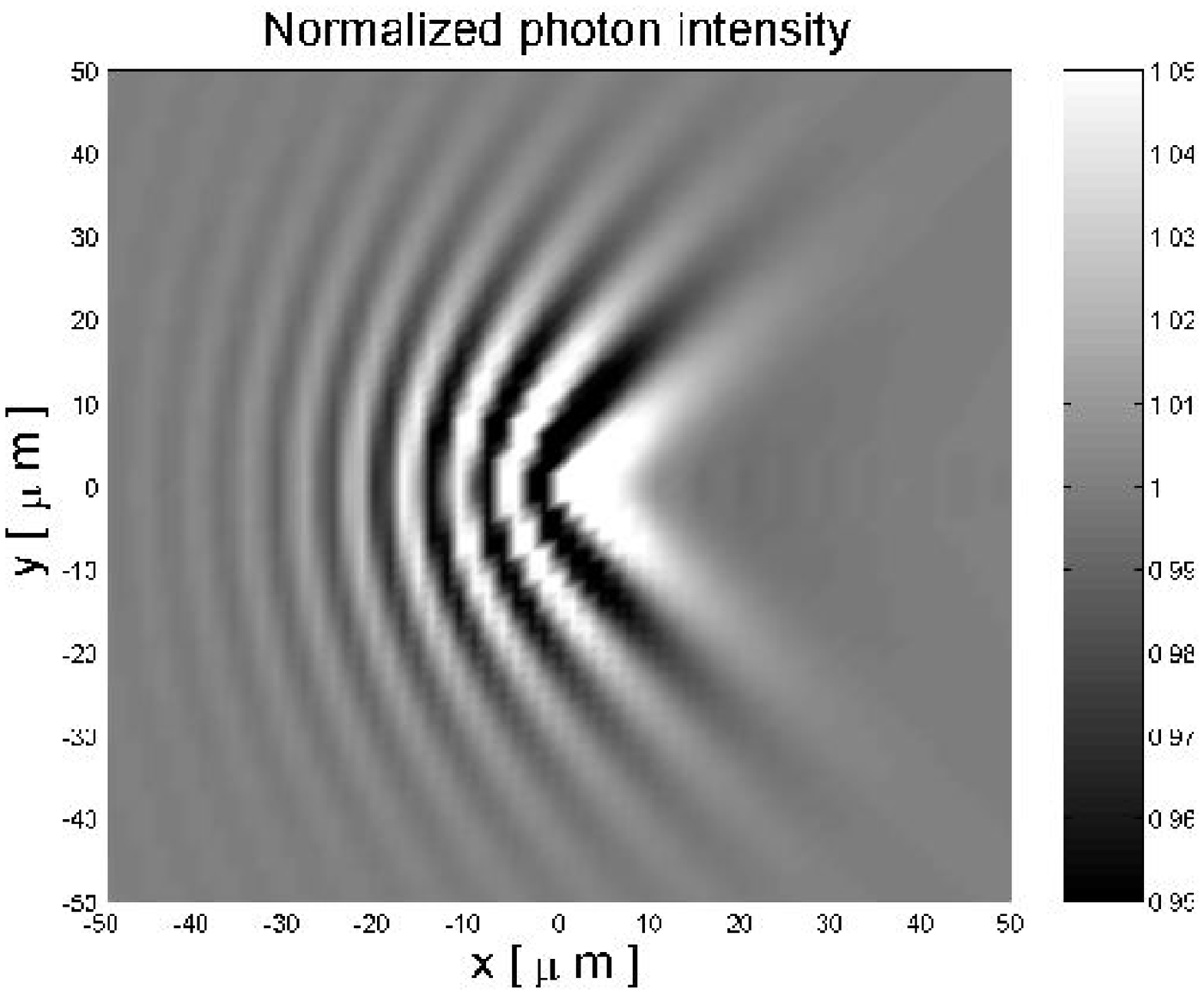}
\caption{ Spatial profile of the normalized cavity photon density.
Same parameters as in Fig. \ref{fig_gapRRS.eps}. The gray scale of
this plot is saturated in the region around the point defect.
\label{fig_gapRRS_space.eps}}
\end{center}
\end{figure}
\begin{figure}[t]
\begin{center}
\includegraphics[width=6cm,clip]{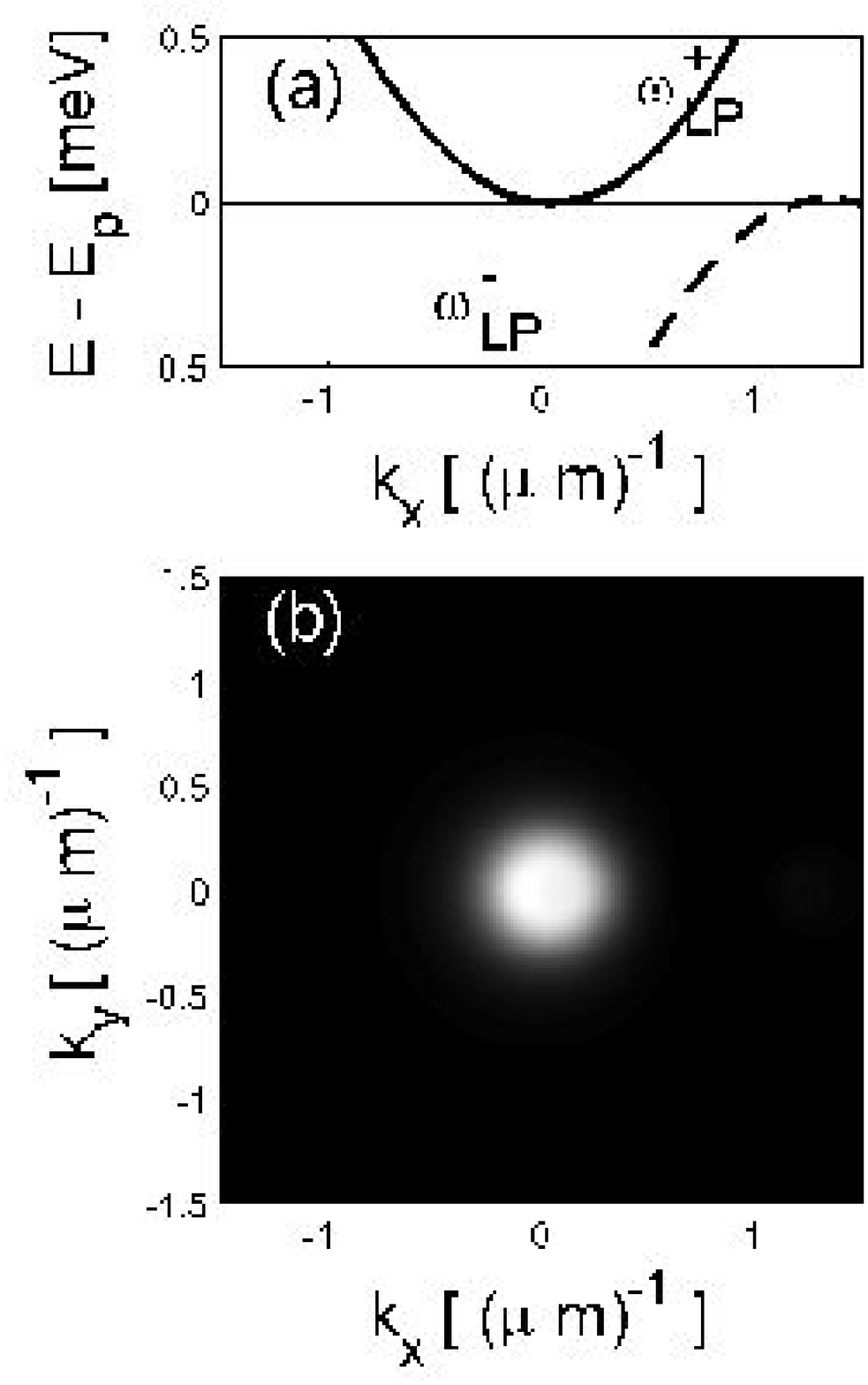}
\caption{Parameters: $\omega_X = \omega_C(0)$, $k_p= 0.7
\mu\textrm{m}^{-1}$ (along the x-axis), $\hbar \omega_p - \hbar
\omega_{LP}(\kk_p) = 0.2\,\textrm{meV}$, $ \hbar g|\Psi^{ss}_X|^2
= 0.6 \textrm{meV}$. \label{fig_biggapRRS.eps}}
\end{center}
\end{figure}
\begin{figure}[h]
\begin{center}
\includegraphics[width=7cm,clip]{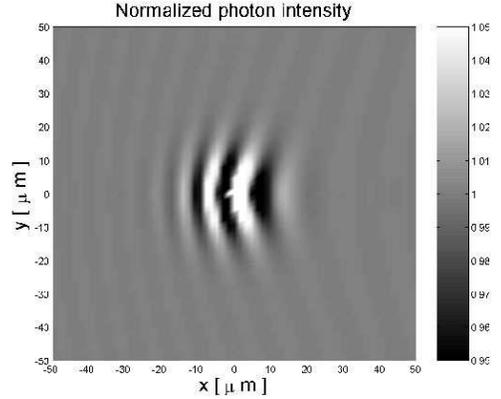}
\caption{ Spatial profile of the normalized cavity photon density.
Same parameters as in Fig. \ref{fig_biggapRRS.eps}. The gray scale
of this plot is saturated in the region around the point defect.
\label{fig_biggapRRS_space.eps}}
\end{center}
\end{figure}
\subsection{Cherenkov regime}

Here, we consider the opposite case $\Delta_p\leq 0$, but with an
excitation density which is not high enough to enter the
superfluid regime characterized by the condition in Eq.
(\ref{condition}). For the sake of clarity, as we have already
done for Fig. \ref{fig_2points.eps} and Fig.
\ref{fig_2points_space.eps}, we take here $V_X=0$, so to
concentrate on the effect of a single defect. This situation can
be realistically realized, e.g., when there is a single photonic
defect (natural or artificial) which is dominant over the
background excitonic disorder. In Fig.
\ref{fig_CherenkovRRS.eps}(a), we consider the resonant case
($\Delta_p = 0$), with a polariton sound speed $c_s < v_p$. As
shown in Fig. \ref{fig_CherenkovRRS.eps}(b) the weak excitation
elastic RRS ring is replaced by an asymmetric pattern, which is
strongly deformed and shows a singularity at the pump wavevector.
The aperture angle $2\theta$ of the singularity at $\kk_p$
satisfies the simple condition $\cos\theta=c_s/v_p$. In this
$c_s<v_p$ regime where the polariton fluid is moving at a
supersonic speed, the defect produces a peculiar real-space
pattern (Fig. \ref{fig_CherenkovRRS_space.eps}) showing linear
Cherenkov-like wavefronts~\cite{Jelley,Cherenkov}. The aperture
$2\phi$ of the Cherenkov angle has the usual value
$\sin\phi=c_s/v_p$. This behavior is easily understood from a
physical standpoint as follows: a moving fluid propagating along
the positive $x$ direction in the presence of a static defect is
equivalent, under a Galilean transformation, to a defect moving in
the negative $x$ direction in a fluid at rest. This situation is a
familiar one, as it corresponds to the wavefronts created by a
moving duck on the surface of a lake. The rounded region on the
left-hand side of the $\kk$-space pattern in Fig.
\ref{fig_CherenkovRRS.eps}(b) was not present in the standard
theory of Cherenkov emission in non-dispersive media~\cite{Jelley}
and it is due to the fact that the Bogoliubov dispersion is linear
only in the neighborhood of $\kk_p$ and then it bends upwards. A
remarkable consequence of this property is the oscillatory
perturbation shown by the real-space pattern upstream with respect
to the defect, as apparent in
Fig.\ref{fig_CherenkovRRS_space.eps}.

In the case $\Delta_p<0$, the branches $\omega_{LP}^+$ and
$\omega_{LP}^-$ are disconnected, as well as the corresponding RRS
emission pattern, as depicted in Fig. \ref{fig_gapRRS.eps}. The
real-space wavefronts shown in Fig. \ref{fig_gapRRS_space.eps} are
still Cherenkov-like, since the separation between the two
branches is relatively small.

The situation of Fig. \ref{fig_biggapRRS.eps} and
\ref{fig_biggapRRS_space.eps} is instead different, with a full
gap opened between the two LP branches. The $\kk$-space emission
is then concentrated around the point $\kk_a$ where the Bogoliubov
dispersion touches the line $\omega = \omega_p$. Correspondingly,
the near-field pattern shows a localized perturbation around the
defect, with a peculiar stripe pattern. This pattern is due to the
interference of the Rayleigh scattering at $\kk_a$ and the pump
beam at $\kk_p$, so that the wavevector corresponding to the
fringes is equal to $\kk_p-\kk_a$.

\section{Conclusions}
In conclusion, in this Festschrift paper, we have presented a
comprehensive analysis of the exotic collective excitations of a
moving polariton fluid driven by a continuous-wave optical pump,
which were recently predicted in a short
Letter\cite{PRL_superfluid}. We have analyzed in detail the
interplay between the non-trivial dispersions of the polariton
Bogoliubov excitations and the onset of single-mode (Kerr) or
multi-mode (parametric) instabilities. We have studied the
propagation of the polariton fluid in presence of static
perturbation potential acting both on the photonic and exciton
component of the polariton field. In particular, we have shown the
strict connection between the dispersion of the elementary
excitations in a quantum fluid of microcavity polaritons and the
intensity and shape of the resonant Rayleigh scattering on
defects. We have pointed out some experimentally accessible
consequences of polaritonic superfluidity for realistic
microcavity parameters. In addition, we have shown that in the
present non-equilibrium system, it is possible to have spectra of
collective excitations, which are not accessible in systems close
to the thermal equilibrium, such as superfluid helium or the
ultracold atomic condensates. These spectra of excitations can be
dramatically changed by tuning the pump excitation parameters,
namely its frequency, incidence angle and intensity.  We have
shown a very rich phenomenology in the far-field and near-field
images of the resonant Rayleigh scattering emission.

\section{Acknowledgments}
 LPA-ENS is a "Unit\'e de Recherche de l'Ecole Normale Sup\'erieure et des
Universit\'es Paris 6 et 7, associ\'ees au CNRS". CC wishes to
thank all the authors contributing to the present Festschrift
volume for many stimulating discussions on the physics of
semiconductor microcavities. IC acknowledges stimulating
discussions with C. Tozzo and F. Dalfovo on the subject of
modulational instabilities in quantum fluids.


\begin{thebibliography}{99}
\bibitem{ManyBody} D. Pines and P. Nozieres, {\em The theory of
  quantum liquids} Vols.1 and 2 (Addison-Wesley,  Redwood City,
  1966).
\bibitem{AtomicBEC} L. Pitaevskii and S. Stringari, {\em Bose-Einstein
  condensation} (Oxford University Press, 2003).
\bibitem{Superfluidity} A. J. Leggett, Rev. Mod. Phys. {\bf 71}, S318-S323 (1999).
\bibitem{Weisbuch} C. Weisbuch, M. Nishioka, A. Ishikawa, and Y. Arakawa, Phys. Rev. Lett. {\bf
69}, 3314 (1992).
\bibitem{Review} For a recent review, see
Semicond. Sci. Technol. {\bf 18}, {\it Special Issue on
Microcavities}, Guest Editors J. Baumberg and L. Vi\~{n}a,
Publisher S. Quin (Bristol, UK, 2003).
\bibitem{Savvidis} P. G. Savvidis, J. J. Baumberg, R. M. Stevenson, M. S. Skolnick,
D. M. Whittaker, J. S. Roberts, Phys. Rev. Lett. {\bf 84}, 1547
(2000).
\bibitem{Ciuti} C. Ciuti, P. Schwendimann, B. Deveaud, and
  A. Quattropani, Phys. Rev. B {\bf 62}, R4825 (2000).
\bibitem{Saba} M. Saba, C. Ciuti, J. Bloch, V. Thierry-Mieg, R. Andr\'{e}, Le Si Dang,
S. Kundermann, A. Mura, G. Bongiovanni, J. L. Staehli, B. Deveaud,
Nature (London) {\bf 414}, 731 (2001).
\bibitem{offbranch} P. G. Savvidis, C. Ciuti, J. J. Baumberg, D. M. Whittaker, M.
S. Skolnick, and J. S. Roberts Phys. Rev. B 64, 075311 (2001)
\bibitem{Huynh} A. Huynh, J. Tignon, O. Larsson, Ph. Roussignol,
C. Delalande, R. Andr\'{e}, R. Romestain, Le Si Dang, Phys. Rev.
Lett. {\bf 90}, 106401 (2003).
\bibitem{Kundermann} S. Kundermann, M. Saba, C. Ciuti, T. Guillet, U. Oesterle,
J. L. Staehli, and B. Deveaud, Phys. Rev. Lett. {\bf 91}, 107402
(2003).
\bibitem{Baumberg} J. J. Baumberg, P. G. Savvidis, R. M. Stevenson,
A. I. Tartakovskii, M. S. Skolnick, D. M. Whittaker, and J. S.
Roberts, Phys. Rev. B {\bf 62}, R16247 (2000).
\bibitem{Stevenson} R. M. Stevenson, V. N. Astratov, M. S. Skolnick, D. M. Whittaker,
M. Emam-Ismail, A. I. Tartakovskii, P. G. Savvidis, J. J.
Baumberg, and J. S. Roberts, Phys. Rev. Lett. {\bf 85}, 3680
(2000).
\bibitem{HoudreRRSNLin} R. Houdr\'e, C. Weisbuch, R. P. Stanley, U. Oesterle, and M. Ilegems, Phys. Rev. Lett. {\bf 85}, 2793 (2000).
\bibitem{Whittaker} D. M. Whittaker, Phys. Rev. B {\bf 63}, 193305 (2001)
\bibitem{Messin} G. Messin, J. Ph. Karr, A. Baas, G. Khitrova, R. Houdr\'{e},
R. P. Stanley, U. Oesterle, E. Giacobino, Phys. Rev. Lett. {\bf
87}, 127403 (2001).
\bibitem{Dasbach} G. Dasbach, M. Schwab, M. Bayer, and A. Forchel, Phys. Rev. B {\bf
64}, 201309 (2001).
\bibitem{Dasbach_wire} G. Dasbach, M. Schwab, M. Bayer, D.N. Krizhanovskii, A. Forchel,
Phys. Rev. B {\bf 66}, 201201 (2002).
\bibitem{Ciuti_01} C. Ciuti, P. Schwendimann, A. Quattropani,
Phys. Rev. B {\bf63}, 041303(2001).
\bibitem{Schwendimann} P. Schwendimann, C. Ciuti, A. Quattropani,
Phys. Rev. B {\bf 68}, 165324 (2003).
\bibitem{squeezing}  J. Ph. Karr, A. Baas, R. Houdr\'{e}, Elisabeth Giacobino,
Phys. Rev. A 69, 031802 (2004).
\bibitem{twin} J. Ph. Karr, A. Baas, and E. Giacobino,
Phys. Rev. A {\bf 69}, 063807 (2004).
\bibitem{Ciuti_branch} C. Ciuti, Phys. Rev. B {\bf 64}, 245304
(2004).
\bibitem{Langbein} W. Langbein, Proc. 26th Int. Conf. on the Physics of Semiconductors; ICPS 26 (Edinburgh, UK,
2002).
\bibitem{Langbein_eight} W. Langbein, Phys. Rev. B {\bf 70}, 205301(2004).
\bibitem{Savasta_entangled} S. Savasta, O. Di Stefano, V. Savona,
W. Langbein, cond-mat/0411314.
\bibitem{PRL_superfluid} I. Carusotto and C. Ciuti, Phys. Rev.
Lett. {\bf 93}, 166401 (2004).
\bibitem{RRS}     H. Stolz, D. Schwarze, W. von der Osten, and
  G. Weimann, Phys. Rev. B {\bf 47}, 9669 (1993).
\bibitem{HoudreRRSLin} R. Houdr\'e, C. Weisbuch, R. P. Stanley, U. Oesterle, and M. Ilegems, Phys. Rev. B {\bf 61}, R13333 (2000).
\bibitem{Langbein_ring} W. Langbein and J. M. Hvam, Phys. Rev.
Lett. {\bf 88}, 047401 (2002), and references therein.
\bibitem{Ciuti_Review} C. Ciuti, P. Schwendimann, and A. Quattropani,
  Semicond. Sci. Technol. {\bf 18}, S279-S293 (2003).
\bibitem{savasta} S. Savasta, O. Di Stefano, and R. Girlanda,
  Phys. Rev. Lett. {\bf 90}, 096403 (2003).
\bibitem{NiuModInst} B. Wu and Q. Niu, Phys. Rev. A {\bf 64}, 061603(R)
  (2001).
\bibitem{BistableExp} A. Baas, J. Ph. Karr, H. Eleuch, and E. Giacobino,
Phys. Rev. A {\bf 69}, 023809 (2004).
\bibitem{BistableExp2} N. A. Gippius, S. G. Tikhodeev, V. D. Kulakovskii, D. N. Krizhanovskii, A. I. Tartakovskii,
  Europhys. Lett. {\bf 67}, 997 (2004).
\bibitem{BistableExp3} A. Baas, J. Ph. Karr, M. Romanelli, A. Bramati, and E.
Giacobino, Phys. Rev. B {\bf 70}, 161307 (R) (2004).
\bibitem{Bistability} R. W. Boyd, {\em Nonlinear Optics} (Academic
  Press, London, 1992).
\bibitem{Ciuti_PL} C. Ciuti, P. Schwendimann, A. Quattropani,
Phys. Rev. B {\bf 63}, 041303(R) (2001); D. M. Whittaker, Phys.
Rev. B {\bf 63}, 193305 (2001).
\bibitem{Jelley} J. V. Jelley, {\em Cerenkov radiation and its
applications}, Pergamon Press, 1958.
\bibitem{Cherenkov}  I. Carusotto,
M. Artoni, G. C. La Rocca, F. Bassani, Phys. Rev. Lett. {\bf 87},
064801 (2001).



\end{thebibliography}
\end{document}